\documentclass[conference]{IEEEtran}

\usepackage{xcolor}
\usepackage{soul, hyperref}
\usepackage{caption}
\usepackage{subcaption}
\usepackage{tcolorbox}
\usepackage[ruled,vlined]{algorithm2e}
\usepackage{url}
\usepackage{amsthm,amsmath}

\newtheorem{definition}{Definition}

\usepackage{multirow}
\usepackage{colortbl}

\DeclareMathOperator*{\argmin}{arg\,min}

\newcommand{\ourapproach}{\textit{SnatchML}}

\ifCLASSINFOpdf
\else
\fi
\begin{document}
\title{SnatchML: Hijacking ML models without Training Access}

% % author names and affiliations
% % use a multiple column layout for up to three different
% % affiliations
\author{\IEEEauthorblockN{Mahmoud Ghorbel}
\IEEEauthorblockA{IEMN CNRS-8520, UPHF, \\
INSA Hauts-de-France}
\and
\IEEEauthorblockN{Halima Bouzidi}
\IEEEauthorblockA{Univ. of California, \\ Irvine, US}
\and%  
\IEEEauthorblockN{Ioan Marius Bilasco}
\IEEEauthorblockA{Univ. Lille, CNRS, Centrale Lille, \\
UMR 9189 CRIStAL, France}
\and
\IEEEauthorblockN{Ihsen Alouani}
\IEEEauthorblockA{IEMN CNRS-8520, UPHF, \\ INSA Hauts-de-France\\ CSIT, Queen's University Belfast, UK}
}

\maketitle

\begin{abstract}
% \begin{abstract}
The widespread deployment of Machine Learning (ML) models has been accompanied by the emergence of various attacks that threaten their trustworthiness and raise ethical and societal concerns. One such attack is model hijacking, where an adversary seeks to repurpose a victim model to perform a different task than originally intended. Model hijacking can cause significant accountability and security risks since the owner of a hijacked model can be framed for having their model offer illegal or unethical services. 
Prior works consider model hijacking as a training time attack, whereby an adversary requires full access to the ML model training. 
In this paper, we consider a stronger threat model for an inference-time hijacking attack, where the adversary has no access to the training phase of the victim model. 
Our intuition is that ML models, which are typically over-parameterized, might have the capacity to (unintentionally) learn more than the intended task they are trained for. 
We propose \ourapproach, a new training-free model hijacking attack, that leverages the extra capacity learnt by the victim model to infer different tasks that can be semantically related or unrelated to the original one. 
Our results on models deployed on AWS Sagemaker showed that \ourapproach~can deliver high accuracy on hijacking tasks. Interestingly, while all previous approaches are limited by the number of classes in the benign task, \ourapproach~ can hijack models for tasks that contain more classes than the original.
We explore different methods to mitigate this risk; We propose meta-unlearning, which is designed to help the model unlearn a potentially malicious task while training for the original task. We also provide insights on \textit{over-parametrization} as a possible inherent factor that facilitates model hijacking, and accordingly, we propose a compression-based countermeasure to counteract this attack. We believe this work offers a previously overlooked perspective on model hijacking attacks, presenting a stronger threat model and higher applicability in real-world contexts. Our code is available at \url{https://github.com/ihsenLab/SnatchML}. %\textcolor{red}{Halima: The abstract is very long, shrinking is needed.}

% \href{https://drive.google.com/drive/folders/1slPjYsFR_nV5iZmK3pj7MEEAc8zMkazY?usp=sharing}{\hl{Link to the Google Drive Folder}} \footnote{https://aws.amazon.com/sagemaker/}

% \end{abstract}    

\end{abstract}


% \fi
%%\IEEEpeerreviewmaketitle

\section{Introduction}
\label{sec:intro}

Machine Learning models have demonstrated cutting-edge performance across a broad spectrum of applications, progressively expanding to be deployed into domains with security-critical and privacy-sensitive implications, such as healthcare, financial sectors, transportation systems, and surveillance. However, as the massive adoption of ML models continues to rise, a variety of attacks with different threat models have emerged, which can jeopardize ML models' trustworthiness. For example, adversarial attacks ~\cite{carlini2017towards,huang2011adversarial,goodfellow2014explaining, neuroattack, dap} are popular inference-time attacks that compromise the security of the model by causing it to misclassify to the attacker's advantage. %Recently, model hijacking attacks have been proposed as is recent training time threat model has also emer, attacks like  also emerged as threats to
The necessity of large amounts of data and high computational resources at the training stage has introduced another attack surface against ML, where the adversary interferes with the model training. Such attacks are usually referred to as training time attacks. Within this category, backdoor and data poisoning attacks are two of the most popular ones \cite{sun2019can,bagdasaryan2020backdoor,biggio2012poisoning,naseri2020toward}.

\noindent\textbf{Model hijacking.} Recently, a new type of training time attacks known as model hijacking has been introduced \cite{usenix_hijack, salem2021get, elsayed2018adversarial, mallya2018packnet}.
In model hijacking attacks, the adversary's objective is to take control of a target model and repurpose it to perform a completely different task, referred to as \textit{hijacking task} without the model owner's knowledge or consent. This can cause accountability risks for the model owner; they can be framed to be offering illegal or unethical services. For example, an adversary can repurpose a benign image classifier into a facial recognition model for illegal surveillance (e.g. on data powered by IoT bots) or for NSFW material. Other possible unethical or illegal scenarios include hijacking a benign model to enable systematic discrimination based on gender, race or age.  Among first works on this threat, Salem et al.~\cite{salem2021get} proposed Model Hijacking attack that hides a model covertly while training a victim model. In the same direction, Mallya et al.~\cite{mallya2018packnet} proposed Packnet that trains the model with multiple tasks.
%The conventional execution of these attacks is concretely performed through data poisoning aimed to repurpose the victim model trained for an \textit{original task} to be able to perform a hijacking task without reducing the utility on the original task. 

 The threat model of these attacks is similar to data poisoning/backdoor attacks, i.e., it requires access to the training process of the victim model. However, in this work, we consider an more critical threat model where the attacker has limited access capabilities. Specifically, the adversary has no access to the training data or process and has access to the model \textit{at inference time only} after deployment. 
 Moreover,  while existing hijacking attacks focus on scenarios where the attacker aims to frame the model owner, we also consider a different setting, where the model owner intentionally misuses a benign/compliant model for malicious activities in a covert manner. In this case, they may avoid altering the input data to evade detection. For instance, consider a scenario where a CCTV camera is used by authorities or a company for a legitimate purpose, such as emotion recognition, utilizing a model from a trusted vendor. With \ourapproach, the same model can be covertly repurposed on the same input stream for malicious activities, such as tracking minorities.

\begin{figure}[htp]
  \centering
  \includegraphics[width=1\columnwidth]{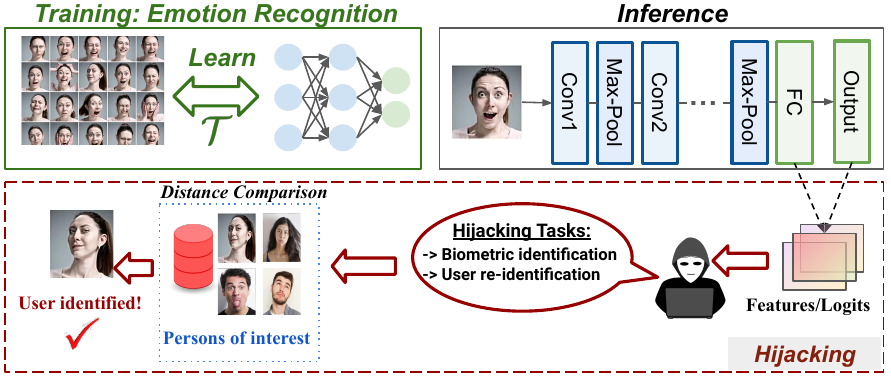}
  \caption{An attacker can use the output of a model trained on task $\mathcal{T}$ to infer a different task ($\mathcal{T}'$). For example, a user can be identified by simply comparing the similarity between feature maps of a query and a reference identity class.} 
  \label{fig:attack}
\end{figure}

\ourapproach~ leverages the capacity of benign ML models trained on clean datasets to acquire "extra-knowledge" to infer a different (potentially malicious) task. The adversary, having access to the deployed model, utilizes "benign extracted knowledge" to infer the hijacking task. Specifically, we analyze the use of either logits (in a black-box scenario) or feature vectors (in a white-box scenario) to classify unknown input samples, using distance measures in the latent space, to previously known samples associated with the classes relevant to the hijacking task. The proposed approach is detailed in Section \ref{sec:approach}.  Figure \ref{fig:attack} gives a high-level illustration of the attack setting in a scenario where the original task is emotion recognition and the hijacking task is biometric identification.

% Announce the second (more general) case, i.e., random projection
To demonstrate our attack methodology, we conduct an analysis across various scenarios with pretrained ML models deployed on AWS. This analysis initially focuses on hijacking tasks that are semantically related to the original task. Specifically, we examine four original tasks where we consider a pre-trained model and demonstrate that, in each case, an attacker with restricted access can exploit a pre-trained model for a hijacking task that shares semantic overlap with the original task (Section \ref{sec:scenario1ER}, \ref{sec:scenario5_age}, \ref{sec:scenario3_Pneu} and \ref{sec:scenario4_ECG}). %For instance, in the first scenario (detailed in Section \ref{sec:scenario1ER}), we investigate models developed for emotion recognition (ER). Interestingly, we find that ER models can be hijacked for biometric identification purposes by merely accessing the logits, without the need to access the model training. 
%Additionally, we present other cases where the attacker hijacks a Pneumonia Diagnosis model (Section \ref{sec:scenario3_Pneu}) to classify viral/bacterial origin, and an age estimation model for recognizing both gender and ethnicity (Section \ref{sec:scenario5_age}).
%\hl{we need to mention the different modality (and maybe cross-modality if the experiments work)--}
While these attacks illustrate the hijacking risk under a stronger threat model than what is currently established in literature, the prerequisite of relatedness between the original and hijacking tasks may limit the scope and impact of such attack strategy. Therefore, in Section \ref{sec:unrelated}, we investigate the general case where the relatedness constraint is relaxed. Surprisingly, we found that \ourapproach~ is capable of hijacking a deployed model for a task that is totally unrelated to the original one. We attempt to provide an explanation of these findings in Section \ref{sec:explain}. Our hypothesis is that the overparametrization of ML models is a core reason behind this phenomenon; \textbf{(i)} it provides capacity to learn clues useful for related tasks, and \textbf{(ii)} it results in a hyper-dimensional representation of data in the latent space, which acts akin to a random projection block, enabling the inference of unrelated tasks.

% does not ensure it will be immune to repurposing for unethical or illegitimate activities. Therefore, additional safeguards and assurances are necessary.
% We believe this work has implications on risk-based AI regulatory approaches. In fact, debates regarding regulating AI-powered systems are drawing increasing attention around the world. Particularly, the European Union (EU) proposed risk-based approach from the European Commission has been particularly prominent in this regard. 

We  propose two methods to mitigate \ourapproach's risk: We first propose \textit{meta-unlearning}, which helps unlearning the potentially malicious task while learning the original task. The second defense method is based on our study of the \textit{over-parametrization}'s impact on enabling model hijacking; We formulate an optimisation problem to find the most compact model that preserves the accuracy of the original task while being less prone to hijacking attacks.

\noindent \textbf{Contributions.} In summary, our contributions are as follows:

\begin{itemize}
    \item We propose an inference-time model hijacking attack that does not require training access and has no restrictions on the number of classes for the hijacking task, applicable in both black-box and white-box settings. %investigate the risk of ML hijacking attacks with a strong threat model where the attacker does not have access to the training data/process.% Specifically, we propose \ourapproach~ to exploit the unintentionally learnt capabilities of the model. %We make our code (anonymously) available\footnote{Code:  \url{https://anonymous.4open.science/r/SntachMLNDSS-6D47}}.  %Our approach can operate under both black-box and white-box settings.% and we show that an attacker with only access to the model's inference can hijack the model for a different (potentially malicious) task. 
    
    \item We illustrate our study with practical scenarios and show that models trained for benign tasks can be hijacked for unethical use such as biometric identification, gender or ethnicity recognition. Surprisingly, we also find that it is possible to infer hijacking tasks that are totally unrelated to the original task. 
    
    % \item We investigate the over-parametrization as a potential reason behind the capacity of ML models to break the least privilege principle at training. 
    
    \item We investigate the over-parametrization as a potential reason behind the vulnerability to inference-time hijacking, and we propose two approaches to limit the risk of \ourapproach: (i) Meta-unlearning, a novel meta-learning based approach that helps the model unlearn an identified malicious task when training on the original one, and (ii) Model compression to strictly limit the capacity of the model to the original task. 
\end{itemize}

% \hl{%- hijacking sucess = f(nbr of benign task classes)\\
% - hijacking sucess = f(ratio original/hijacking nbr of classes\\
% %- explore as f(classes distinguishability in the hijacking task)\\
% - Upper bound hijacking\\
% - other modalities + cross modality\\
% %- post-softmax\\
% - Top-k classes \\
% - In countermeasures, emphasize the fact that compression doesn't need prior information about hijacking, in contrast to unlearning.}

\noindent
\textbf{Disclosures and ethics.} We disclosed our findings to Amazon AWS. We tested our experiments on AWS SageMaker \cite{sagemaker}, specifically using JumpStart, and were \textit{limited to our own models}. No public or commercial MLaaS systems were impacted.

% We believe this work provides an illustration of new risks to ML models. We hope it will draw attention to the least privilege as a core cybersecurity principle that needs to be taken into account for the training and deployment of trustworthy ML models. 

%\input{sec/02_intuition}
\section{Threat Model} \label{sec:threat}

We consider an attacker who deploys a pre-trained ML model acquired from a trusted third party. The model is securely trained on clean data for a benign original task.

\noindent \textbf{Attacker's objective:} This threat model focuses on scenarios where the attacker aims to re-purpose the deployed ML model for eventual malicious tasks different from the original one. We consider 2 scenarios: The first corresponds to the classical hijacking where the attacker aims to frame the victim, who is the model owner. The second scenario is when the attacker is the model owner who purposefully misuses a benign/compliant model for malicious activities in a covert manner, i.e., without changing the model or the input.
%For example, an attacker wants to hijack a \textit{compliant} pretrained model for unethical or illegal activities without having access to the training. 

% Specifically, the attack involves leveraging potential extra capabilities the model \textit{unintentionally} over-learned about the training data sampled in the training phase that can be maliciously exploited for other tasks not anticipated by the vendor or authorized by the user. Such extra capabilities can manifest as memorizing private information about the users who have been part of the training dataset or learning properties of the legitimate training dataset (e.g., male and female ratio). The attacker, therefore, wants to use a \textit{compliant} pretrained model without extra training or fine-tuning to perform malicious activity. 

%the learned capabilities of the model for tasks not anticipated by the vendor or authorized by the user. The attacker tries to leverage the potential extra-capabilities the model \textbf{unintentionally} learned during the training phase. %We focus specifically on exploiting the knowledge learnt by the model to infer private information.  %The attacker's primary goal is to repurpose the benign model for a task different from its original intent, potentially with malicious intent. Specifically, the exploitation involves leveraging the learned capabilities of the model for tasks not anticipated by the vendor or authorized by the user.

\noindent \textbf{Attacker's Capabilities:} \textit{(i) At training time}, We assume that the model is trained to perform a benign task and that the attacker cannot interfere by any means with the model's training phase, i.e., the adversary does not have the ability to poison the target model’s training dataset, in contradiction with existing hijacking \cite{salem2021get, usenix_hijack} or poisoning \cite{biggio2012poisoning} attacks' assumptions. 

\textit{(ii) At inference time}, we consider that the attacker can have access to the model under 2 different settings: \\ \noindent\textbf{(A) Black-box}: the attacker has access to the output logits of the pre-trained model, but not to its internal architecture. This setting is similar to the conventional hijacking attacks and corresponds to the case where the model owner is the victim of the hijacking, e.g., the victim acquires a model from a trusted third-party and deploys it as an ML-as-a-Service that can be queried through APIs that provides output logits.  \\ \noindent\textbf{(B) White-box}: in this case the attacker can have access to the internal state of the model (e.g., feature maps). This case corresponds to the case where the model owner is the attacker; they can show the ML model's compliance to the regulations by referring to the trustworthy vendor, but use the model for potentially unethical tasks.
% In both cases, we consider that the attacker cannot edit the input

% necessarily to the raw input, perhaps because of a multi-party deployment paradigm. In such settings, raw inputs can be encrypted or masked and only the output of the pretrained model is made available. 

% Given the aforementioned assumptions, we consider a \textit{black-box} setting where the attacker can only access the outputs of the pretrained model (e.g., output feature maps of the model's classifier layers or logits vector).  
Moreover, we assume the attacker has access to a dataset related to the hijacking task. We also assume that this dataset may not be sufficient to train or fine-tune a model, with the sole constraint being that it contains at least one sample for each class involved in the hijacking task.

\section{S\MakeLowercase{natch}ML: General Approach} \label{sec:approach}

\noindent \textbf{Problem Statement--}
Given a model $h_{\theta}(\cdot)$ with parameters $\theta$, securely trained on a clean data distribution $D(X, Y)$ using a Loss function $\mathcal{L}$ to perform a task $\mathcal{T}$. 
The objective is to leverage the hidden capabilities that the victim model $h_{\theta}(\cdot)$ acquired through the design and training process to exploit the model to the adversary's advantage post-deployment. Specifically, the adversary aims to leverage $h_{\theta}(\cdot)$'s \textbf{unintentionally} learnt information/capabilities to hijack it for another task $\mathcal{T'}$.

%Specifically, we want to investigate the following questions:
% The objective is to determine whether the model $h_{\theta}(\cdot)$ has hidden capabilities learned throughout this training process that can be exploited through post-processing at inference time after the model is deployed. Specifically,  we want to investigate the following: 

% \begin{itemize}
%     \item \textbf{Q1--} Can $h_{\theta}(\cdot)$ \textbf{unintentionally} learn information that allow an attacker to hijack it for another task $\mathcal{T'} \neq \mathcal{T}$? 
%     \item \textbf{Q2--} Ultimately, can $h_{\theta}(\cdot)$ still be used to infer a task $\mathcal{T'}$ that is totally unrelated to $\mathcal{T}$? 
% \end{itemize}

% To investigate Q1 and Q2, we propose to use information extracted from $h_{\theta}$ \textbf{at inference time} and without direct access to the raw input and try to exfiltrate information or repurpose the model for a different task. 

\noindent \textbf{Proposed approach--}
Without loss of generality, let's suppose that the model $h(\cdot)$ is trained to perform an original task $\mathcal{T}$ consisting of a multi-class classification with $n$ classes. Given the threat model in Section \ref{sec:threat}, the attacker wants to use the model for another (hijacking) task $\mathcal{T'}$. 

\begin{definition}\label{def:bnf} -- \textbf{(Benign Extracted Knowledge). } We define the benign extracted knowledge facts (BEK) as metadata learnt by a benign model from clean input. For a given model $h(\cdot)$ and an input sample $x$, we note $\zeta_h(\cdot)$ an operator that extracts BEK.
\end{definition}
    
Given Definition \ref{def:bnf}, $\zeta_h(x)$ may correspond in the black-box scenario to  $\zeta_(x)= Z_{h}(x)$, where $Z_{h}= \{ z_i\}_{i=1..n}$ is model's output logits vector, or in the white-box scenario to the learnt features tensor, i.e., $\zeta_h(x)= h_k(x)$, where $h_k(x)$ is the output of the $k^{th}$ layer of the model $h(\cdot)$. 

%The model's re-purposing is performed through post-processing BEN. 
The attacker wants to repurpose $h(\cdot)$ for a hijacking $m$-class classification task $\mathcal{T}'$ through a $0$-shot inference on $\zeta_h(\cdot)$. The attacker has access to a dataset $\mathcal{D}^*={(x^*_i,\ell^*_i), i \in [1,m]}$, which contains only $m$ data samples, each corresponding to a class of the hijacking task $\mathcal{T}'$. 
We propose a straightforward exploitation method that is based on the distance in the BEK space. Given an input sample $x_s$ and the corresponding BEK vector $\zeta_h(x_s)$ after inference with $h(\cdot)$, the classification of $x_s$ for $\mathcal{T}'$ is inferred as follows: 
\begin{align} \label{eq:apprch}
     & y_{\mathcal{T}'}(x_s) = \ell^*_p, \textit{  s.t.  }  \\ & p =    \underset{i \in [1,m]}{Argmin} \Big[ \delta( \zeta_h(x^*_i), \zeta_h(x_s) ) \Big] 
\end{align}
   Where $\delta$ is a distance metric such as $\ell_2$ or the Cosine Similarity. 

% \ihsen{I need to mention the hyperparameters limitation here } 
In essence, \ourapproach~ takes advantage of the model's capability to distinguish patterns in a data distribution corresponding to a task $\mathcal{T}'$ that was not part of its training, effectively hijacking the model at inference time. SnatchML can be viewed as a form of zero-shot transfer learning, where the victim model, kept entirely frozen, is repurposed for a new unrelated task, in a black-box or white-box setting.

\noindent\textbf{Remark.} It is worth noting that in the proposed approach, there are no a priori constraints on $m$. In fact, in the state-of-the-art attacks, such as \cite{salem2021get}, the number of classes that can be inferred through the hijacking task cannot exceed those in the original task, i.e., $m \leq n$. Therefore, our approach not only assumes a stronger threat model but also overcomes this limitation, enabling the attacker to hijack an $n$-class model for an $m$-class task where $m>n$. We illustrate this case in Section \ref{sec:scenario1ER}. Moreover, since there is no interference with the model during training, the same model can be hijacked to execute more than one task at the same time. This case is illustrated in Section \ref{sec:scenario5_age}.

\section{Scenario 1: Emotion Recognition}\label{sec:scenario1ER}

%\subsection{Context}\label{sec:er_context}
%The original task $\mathcal{T}$ is Emotion Recognition (ER). 
We investigate the vulnerability of a model trained for emotion recognition (ER) to being hijacked at inference time for biometric identification. In this scenario, the adversary has access to a dataset of 'persons of interest' (the hijacking dataset) and aims to identify these individuals by exploiting the original model.

\subsection{Setup} \label{subsec:setup1}

%\noindent\textbf{Datasets. } To run our experiments, we need datasets that are labeled for both the original and the hijacking tasks. For this experiment, we use $4$ different datasets from both real and synthetic distributions:

\noindent\textbf{CK+ Dataset.} The Extended Cohn-Kanade (CK+) \cite{lucey2010extended} is a dataset for ER that contains $593$ video clips from $123$ individuals between the ages of $18$ and $50$. These videos capture transitions between neutral and peak emotions: anger, disgust, fear, happiness, sadness, and surprise. 

\noindent\textbf{Olivetti Faces Dataset. \cite{samaria1994parameterisation}}  is a facial recognition dataset comprising $400$ face images gathered from 40 unique individuals. The dataset captures variations in lighting and facial expression with a uniform black background, and labeled with the identities of the $40$ individuals.

\noindent\textbf{Celebrity Dataset. }  We extracted $79$ facial images for $9$ different celebrities. Then, using BRIA AI \cite{bria}, which is based on a visual generative model \cite{elasri2022image}, we generated emotion-specific images from the neutral images. These images are labeled for seven emotions (anger, disgust, fear, happiness, sadness, surprise, and neutral) and associated with $9$ unique individuals used as identity labels. 
%This valuable resource can be utilized by researchers and practitioners in computer vision for further exploration and experimentation.
%The annotation spans emotions (anger, disgust, fear, happiness, sadness, surprise, and neutral), as well as identity labels.
%\textcolor{red}{\hl{Is there a rationale we did not use existing celebrity datasets like CelebA instead of building our own?}} \textcolor{green}{{We collected images of neutral faces seen from the front with an empty background in order to generate images with facial emotions, unlike, for example, CelebA, which depicts non-neutral faces taken from various angles and backgrounds.}} 
% (\url{https://www.kaggle.com/datasets/jessicali9530/celeba-dataset}). \textcolor{blue}{Thanks. That's great --  But is it }

\noindent\textbf{Synthetic Dataset.} To cover the synthetic data distribution case, we create a new dataset labeled for both emotions and identities using MakeHuman \cite{bastioni2008ideas}, an open-source tool designed to generate virtual human faces. This dataset encompasses $395$ images labeled with both emotion and identity. Among these images, $47$ depict neutral expressions, while the remaining $348$ represent six distinct emotions. The dataset contains a total of $47$ individual identities, each associated with varying emotional expressions. %, thereby providing a comprehensive representation of human emotional and identity diversity.

\begin{figure}[tp]
  \centering
  \includegraphics[width=\linewidth]{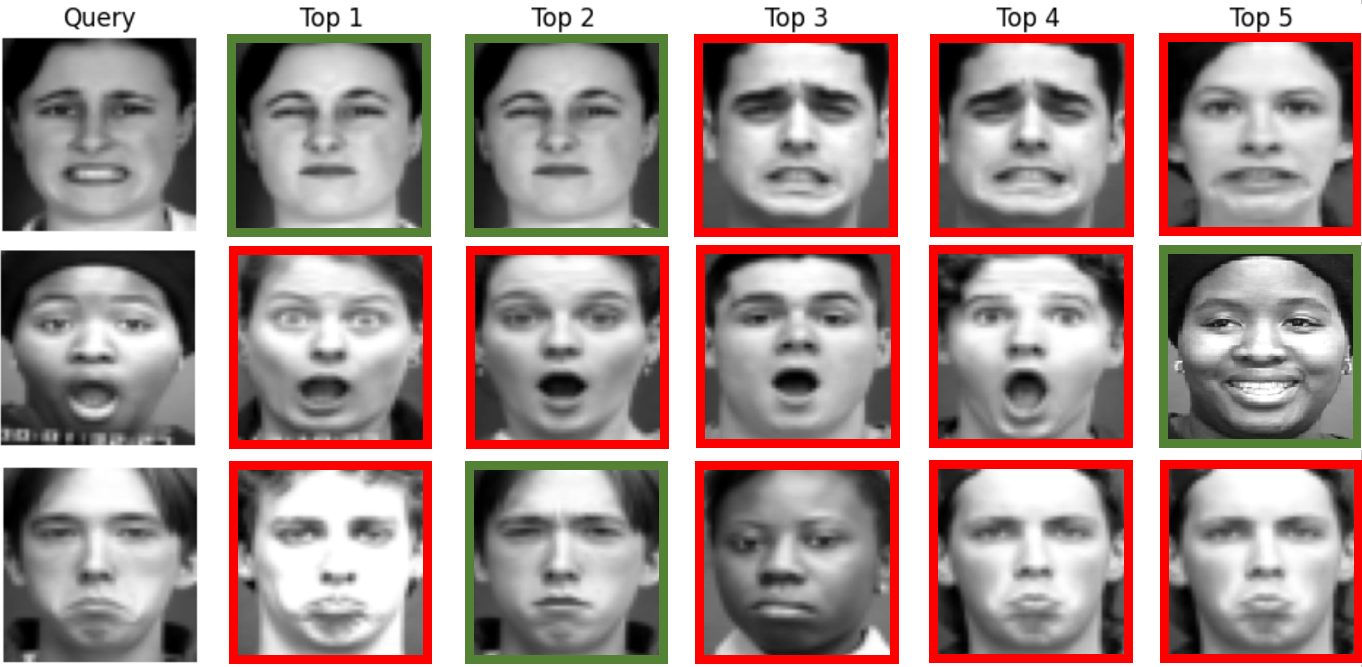}
  \caption{Examples of the hijacked ER model's output top-5 candidates for users re-identification from CK+ dataset.}
  \label{fig:faces-reid}
\end{figure}

\begin{figure}[tp]
  \centering
  \includegraphics[width=\linewidth]{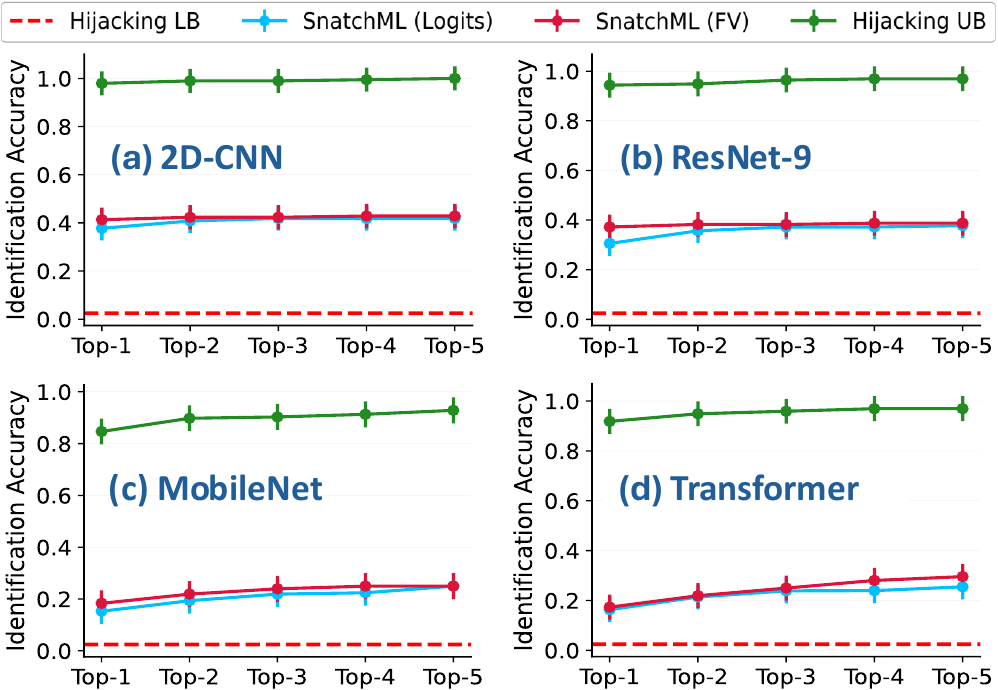}
  \caption{\ourapproach~performance on hijacking an ER model for users re-identification. 'Hijacking LB' and 'Hijacking UB' refer to the lower and upper bounds, respectively. 
  }
  \label{fig:top5-ck+}
\end{figure}

We evaluate this attack scenario by designing an experimental setup with $4$ pre-trained ER models: (\textbf{i}) 2D-CNN \cite{allaert2022comparative, poux2021dynamic} with three convolutional layers, each followed by ReLU activation and max-pooling followed by two fully connected layers, (\textbf{ii}) 
ResNet-9 \cite{he2016deep},  (\textbf{iii}) MobileNet \cite{howard2017mobilenets}, and (\textbf{iv}) Vision Transformer \cite{dosovitskiy2020image}. We train the models for ER on CK+ \cite{lucey2010extended} as the original task $\mathcal{T}$ using a learning rate of $0.001$ and Adam optimizer for $100$ epochs. %We use the CK+ \cite{lucey2010extended} as a training dataset for the ER task where data are split into $70\%$ for training, $10\%$ for validation, and $20\%$ for test.

\noindent\textbf{Comparison.} To evaluate the accuracy of the hijacking tasks, we consider a lower bound and an upper bound (UB): The lower bound (LB) corresponds to the random guessing probability, while the UB corresponds to an \textbf{unconstrained} version of \cite{salem2021get} without covertness requirement, which is practically equivalent to \textit{freely poisoning} the model by training it on both the original and hijacking tasks explicitly. Additionally, we compare with the accuracy achieved when the baseline models are trained specifically for the hijacking task, in scenarios where dataset size and the number of samples per class are not limiting factors.

\subsection{ER Systems for user re-identification} \label{subsec:re-id}
We refer to this scenario as \textit{user re-identification}, as it involves the adversary aiming to identify users whose data (images) were used to train the original ER model.  It is important to note that in our experiments, the hijacking is performed on users' data that is \textbf{not} part of the original training dataset, to follow our threat model.
We use the \textit{top-N} ranked reference users as an evaluation metric for the the hijacking attack performance. %Regarding facial recognition, the Top-N metric is an essential tool for assessing system effectiveness \cite{wang2021deep}. 
% Here's the old justficiation from the last version.
The top-N metric is particularly relevant for the evaluation of identification tasks \cite{wang2021deep}; it can provide a contextual method for narrowing down potential candidate users. Figure \ref{fig:faces-reid} provides an illustration with queries and their corresponding top-5 samples. In the second line of this figure, the query image features a person with dark skin and only one individual with dark skin appears in the top-5 reference users. The attacker could reasonably deduce that the unknown query image likely belongs to this individual. Figure \ref{fig:top5-ck+} shows the top-1 to top-5 accuracy results for the hijacking task, presented as mean and standard deviation after conducting $10$-fold experiments. Considering our hijacking task involves identifying $85$ persons, we use random classification as a reference ($1.1\%$). \ourapproach~achieves over $40\%$ top-1 accuracy for 2D-CNN, while the hijacking UB is up to $98\%$. % \textcolor{blue}{and $99\%$ when the model is trained specifically to the re-identification task}.  
%\ihsen{@me TODO: comment on UB}%  in this context, corresponds to a baseline accuracy of $1.1\%$.
Although all four tested models demonstrate significant performance, the results shown in Figure \ref{fig:top5-ck+} reveal disparities among them, suggesting that different architectures may have varying levels of susceptibility to the proposed model hijacking for the same task. For example, while the attack on 2D-CNN and ResNet-9 succeeded with around $40\%$ accuracy, the ViT and MobileNet were more robust with an average Top-5 accuracy not exceeding $30\%$. 

\subsection{ER systems learn biometric identification} \label{subset:id}
In this second scenario, we consider a setting where the attacker aims to hijack the model for biometric identification of users whose data is not necessarily a part of the ER model's training dataset.  Following the same approach in Section \ref{subsec:re-id}, we exploit the pre-trained ER model to identify unseen users based on the extracted BEK. As for the hijacking reference database, we consider two main settings: \\
\noindent\textbf{(\emph{i})} The attacker has a reference database from a different distribution from the training dataset, comprising distinct users' images with several facial expressions captured at different times and under different lighting conditions.

\noindent\textbf{(\emph{ii})} The attacker has more restricted access, specifically to images of users with a \textbf{neutral facial expression}. These images are typically sourced from official documents such as passports or staff cards. It is important to note that in both scenarios, we assume the users are not part of the training dataset of the ER model.

% \begin{figure}[ht]
%   \centering
%   \includegraphics[width=.9\linewidth]{attack_figures/celebrity_top5.png}
%   \caption{ER model used to to identify users from the Celebrity dataset. The TOP-5 similar users from the hijacking reference database are displayed for each query.}
%   \label{fig:faces-celebrity}
% \end{figure}

\begin{figure}[tp]
  \centering
  \includegraphics[width=\linewidth]{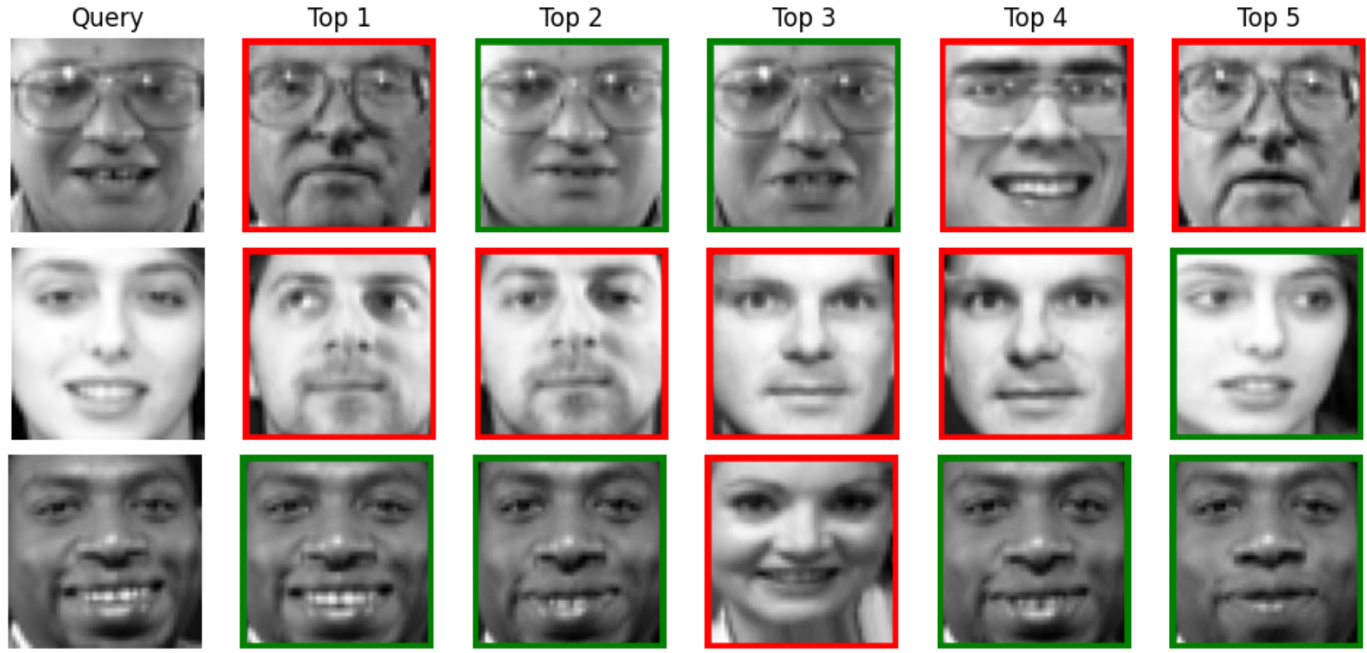}
  \caption{ER model used to to identify users from the Olivetti dataset \cite{samaria1994parameterisation}. The top-5 similar users from the hijacking reference database are displayed for each query.}
  \label{fig:faces-olivetti}
\end{figure}

\begin{figure}[tp]
  \centering
  \includegraphics[width=\linewidth]{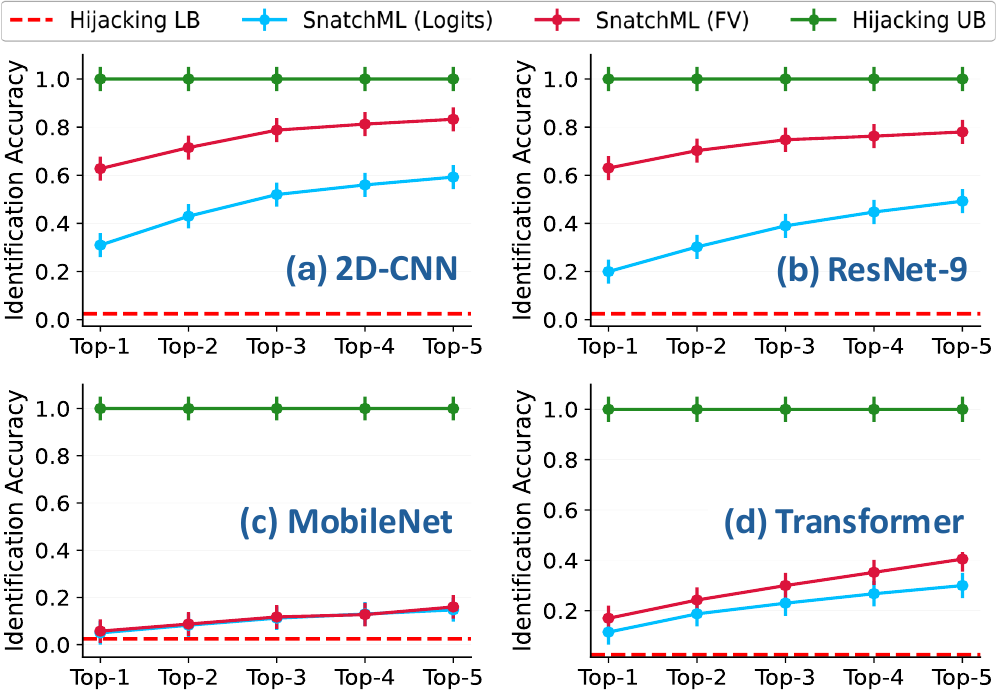}
  \caption{\ourapproach~performance on hijacking an ER model for biometric identification of users from the Olivetti dataset. 
  %\ihsen{\hl{UB here?}} --> Done
  }
  \label{fig:top5-olivetti}
\end{figure}

\noindent \textbf{\underline{Case (\emph{i})}}: To evaluate the identification attack, we use Olivetti dataset \cite{samaria1994parameterisation}. We perform ER queries with images from Olivetti dataset to get the output FV/Logits and run \ourapproach. Figure \ref{fig:faces-olivetti} displays illustration samples of the identification task for Olivetti. An interesting observation is highlighted in the first row, where the top-5 output candidates for a query involving a person wearing glasses also predominantly feature candidates wearing glasses. This indicates that the ER model has learned to recognize this accessory, despite its irrelevance from emotion recognition perspective. The second row in Figure \ref{fig:faces-olivetti} also highlights the significance of the top-5 metric. In this case, the query image corresponds to a female individual with $4$ out of $5$ of the closest candidates representing male subjects, with the first female  candidate appearing only at rank-5, potentially leading to a precise identification. Figure \ref{fig:top5-olivetti} shows the hijacking attack accuracy results on the four pre-trained ER models in terms of top-1 to top-5 accuracy. The attacks demonstrate notable success, with the top-1 accuracy achieving over $60\%$ on 2D-CNN and ResNet-9. Although the lower success rate on MobileNet and the ViT remains consistent in this setting, the hijacking performance is still considerable considering a random guessing probability of $\sim2.5\%$. 

\begin{figure}[h]
  \centering
  \includegraphics[width=\linewidth]{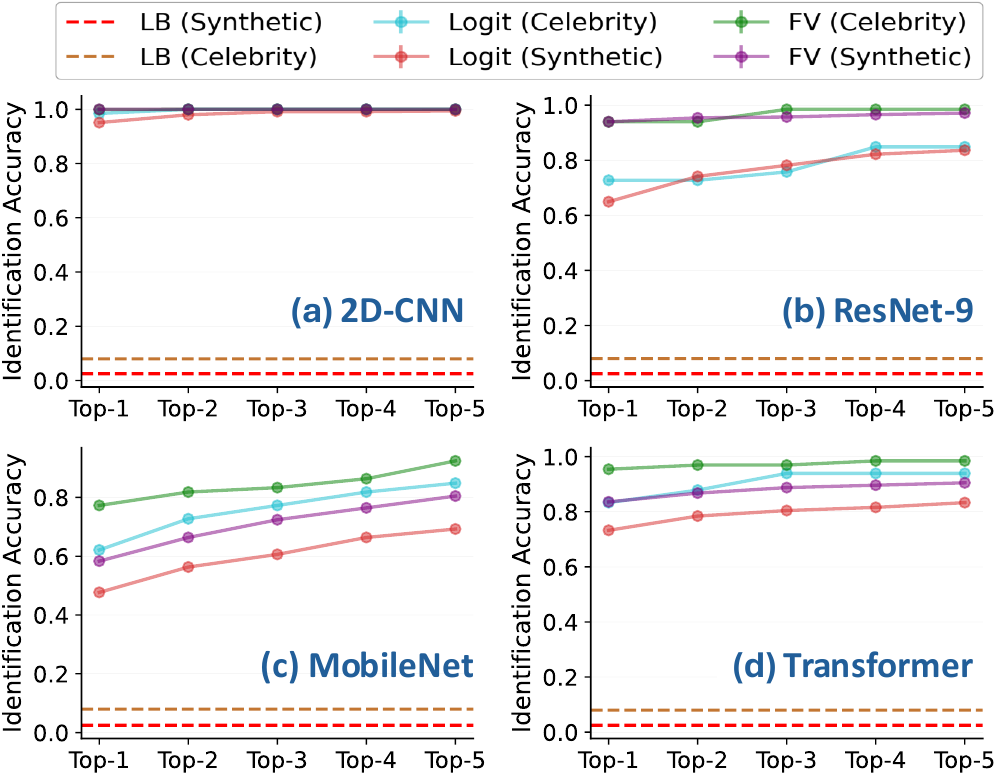}
  \caption{SnatchML Performance in Hijacking an ER Model for Identification Using Real (Celebrity) and Synthetic Datasets.   }
  \label{fig:top5gendatasets}
\end{figure}

\noindent \textbf{\underline{Case (\emph{ii})}}: In this scenario, we assume that the attacker has access only to neutral images of targeted users (e.g. acquired from ID cards). To conduct our attack, we use two datasets featuring individuals with neutral facial expressions, namely Celebrity and Synthetic.  The ER model is pretrained on CK+ dataset. We assess the effectiveness of the identification hijacking attack by performing ER queries with images from both Celebrity and Synthetic datasets. The accuracy of the hijacking task using the four pretrained ER models is shown in Figure \ref{fig:top5gendatasets}. Notably, the top-1 identification accuracy reached up to approximately $100\%$ in the 2D-CNN model under both black-box and white-box attacks for both datasets. The least successful attack was observed with MobileNet under the black-box scenario, where it achieved approximately $48\%$ top-1 identification accuracy with the Synthetic dataset while the random guess probability is $\sim2\%$. %\halima{We need to update this dicussion as we've removed the celebrity figure.}

%\section{Scenario 2: Action Recognition}\label{sec:scenario2AR}\label{sec:scenario2AR}
%\input{sec/4-2_AR}

% \section{Scenario 4: Arrhythmia}\label{sec:scenario4_arryth}
% \input{sec/4-4_ecg}

\section{Scenario 2: Age/Gender/Ethinicity} \label{sec:scenario5_age}

\begin{table*}[htp]
\centering
\caption{\ourapproach~Performance on the UTK dataset. We train 04 benign ML models (2D-CNN, ResNet-9, MobileNet, and ViT) to predict Age, Gender, and Ethnicity. We hijack each model to infer other properties (e.g., from age, we infer gender/ethnicity). }\label{tab:utk_attack}
\scalebox{0.85}{
\begin{tabular}{cc|ccc|ccc|ccc|ccc} 
\hline
\multicolumn{2}{c|}{\textbf{Model}} & \multicolumn{3}{c|}{\textbf{2D-CNN }} & \multicolumn{3}{c|}{\textbf{ResNet-9 }} & \multicolumn{3}{c|}{\textbf{MobileNet }} & \multicolumn{3}{c}{\textbf{Transformer }} \\
\multicolumn{2}{c|}{\textbf{Original~Task }} & \textbf{Age} & \textbf{Gender} & \textbf{Ethnicity} & \textbf{Age} & \textbf{Gender} & \textbf{Ethnicity} & \textbf{Age} & \textbf{Gender} & \textbf{Ethnicity} & \multicolumn{1}{c|}{\textbf{Age}} & \textbf{Gender} & \textbf{Ethnicity} \\
\multicolumn{2}{c|}{\textbf{Original Accuracy}} & {\cellcolor[rgb]{0.984,0.937,0.937}}0.682 & {\cellcolor[rgb]{0.984,0.937,0.937}}0.876 & {\cellcolor[rgb]{0.984,0.937,0.937}}0.762 & {\cellcolor[rgb]{0.984,0.937,0.937}}0.705 & {\cellcolor[rgb]{0.984,0.937,0.937}}0.897 & {\cellcolor[rgb]{0.984,0.937,0.937}}0.785 & {\cellcolor[rgb]{0.984,0.937,0.937}}0.668 & {\cellcolor[rgb]{0.984,0.937,0.937}}0.864 & {\cellcolor[rgb]{0.984,0.937,0.937}}0.726 & \multicolumn{1}{c|}{{\cellcolor[rgb]{0.984,0.937,0.937}}0.635} & {\cellcolor[rgb]{0.984,0.937,0.937}}0.848 & {\cellcolor[rgb]{0.984,0.937,0.937}}0.727 \\ 
\hline
\multirow{4}{*}{\begin{tabular}[c]{@{}c@{}}\textbf{\textbf{Hijacking Age}}\\\textbf{\textbf{Accuracy}}\end{tabular}} & \textbf{\textbf{Hijacking LB}} & \multicolumn{3}{c|}{{\cellcolor[rgb]{0.945,0.98,0.941}}0.166} & \multicolumn{3}{c|}{{\cellcolor[rgb]{0.945,0.98,0.941}}0.166} & \multicolumn{3}{c|}{{\cellcolor[rgb]{0.945,0.98,0.941}}0.166} & \multicolumn{3}{c}{{\cellcolor[rgb]{0.945,0.98,0.941}}0.166} \\
 & \textbf{\textbf{SnatchML (}Logits)} & {\cellcolor[rgb]{0.949,0.957,0.969}}\textbf{-} & {\cellcolor[rgb]{0.945,0.98,0.941}}0.322 & {\cellcolor[rgb]{0.945,0.98,0.941}}0.369 & {\cellcolor[rgb]{0.949,0.957,0.969}}\textbf{-} & {\cellcolor[rgb]{0.882,0.957,0.906}}0.288 & {\cellcolor[rgb]{0.882,0.957,0.906}}0.347 & {\cellcolor[rgb]{0.949,0.957,0.969}}\textbf{-} & {\cellcolor[rgb]{0.882,0.957,0.906}}0.347 & {\cellcolor[rgb]{0.882,0.957,0.906}}0.326 & {\cellcolor[rgb]{0.949,0.957,0.969}}\textbf{-} & {\cellcolor[rgb]{0.882,0.957,0.906}}0.357 & {\cellcolor[rgb]{0.882,0.957,0.906}}0.343 \\
 & \textbf{\textbf{\textbf{\textbf{SnatchML}}}\textbf{~(}FV)} & {\cellcolor[rgb]{0.949,0.957,0.969}}\textbf{-} & {\cellcolor[rgb]{0.776,0.922,0.757}}0.407 & {\cellcolor[rgb]{0.776,0.922,0.757}}0.420 & {\cellcolor[rgb]{0.949,0.957,0.969}}\textbf{-} & {\cellcolor[rgb]{0.776,0.922,0.757}}0.452 & {\cellcolor[rgb]{0.776,0.922,0.757}}0.436 & {\cellcolor[rgb]{0.949,0.957,0.969}}\textbf{-} & {\cellcolor[rgb]{0.776,0.922,0.757}}0.350 & {\cellcolor[rgb]{0.776,0.922,0.757}}0.346 & {\cellcolor[rgb]{0.949,0.957,0.969}}\textbf{-} & {\cellcolor[rgb]{0.776,0.922,0.757}}0.382 & {\cellcolor[rgb]{0.776,0.922,0.757}}0.422 \\
 & \textbf{Hijacking UB} & {\cellcolor[rgb]{0.949,0.957,0.969}}\textbf{-~} & {\cellcolor[rgb]{0.627,0.867,0.592}}0.669 & {\cellcolor[rgb]{0.627,0.867,0.592}}0.668 & {\cellcolor[rgb]{0.949,0.957,0.969}}\textbf{-~} & {\cellcolor[rgb]{0.627,0.867,0.592}}0.669 & {\cellcolor[rgb]{0.627,0.867,0.592}}0.675 & {\cellcolor[rgb]{0.949,0.957,0.969}}\textbf{-~} & {\cellcolor[rgb]{0.627,0.867,0.592}}0.662 & {\cellcolor[rgb]{0.627,0.867,0.592}}0.670 & {\cellcolor[rgb]{0.949,0.957,0.969}}\textbf{-~} & {\cellcolor[rgb]{0.627,0.867,0.592}}0.636 & {\cellcolor[rgb]{0.627,0.867,0.592}}0.615 \\ 
\hline
\multirow{4}{*}{\begin{tabular}[c]{@{}c@{}}\textbf{\textbf{Hijacking Gender}}\\\textbf{\textbf{Accuracy}}\end{tabular}} & \textbf{\textbf{\textbf{\textbf{Hijacking LB}}}} & \multicolumn{3}{c|}{{\cellcolor[rgb]{0.945,0.98,0.941}}0.500} & \multicolumn{3}{c|}{{\cellcolor[rgb]{0.945,0.98,0.941}}0.500} & \multicolumn{3}{c|}{{\cellcolor[rgb]{0.945,0.98,0.941}}0.500} & \multicolumn{3}{c}{{\cellcolor[rgb]{0.945,0.98,0.941}}0.500} \\
 & \textbf{\textbf{\textbf{\textbf{SnatchML}}}\textbf{~(}Logits)} & {\cellcolor[rgb]{0.882,0.957,0.906}}0.612 & {\cellcolor[rgb]{0.949,0.957,0.969}}\textbf{-} & {\cellcolor[rgb]{0.882,0.957,0.906}}0.549 & {\cellcolor[rgb]{0.776,0.922,0.757}}0.547 & {\cellcolor[rgb]{0.949,0.957,0.969}}\textbf{-} & {\cellcolor[rgb]{0.882,0.957,0.906}}0.529 & {\cellcolor[rgb]{0.882,0.957,0.906}}0.561 & {\cellcolor[rgb]{0.949,0.957,0.969}}\textbf{-} & {\cellcolor[rgb]{0.776,0.922,0.757}}0.542 & {\cellcolor[rgb]{0.882,0.957,0.906}}0.578 & {\cellcolor[rgb]{0.949,0.957,0.969}}\textbf{-} & {\cellcolor[rgb]{0.882,0.957,0.906}}0.548 \\
 & \textbf{\textbf{\textbf{\textbf{SnatchML}}}\textbf{~(}FV)} & {\cellcolor[rgb]{0.776,0.922,0.757}}0.626 & {\cellcolor[rgb]{0.949,0.957,0.969}}\textbf{-} & {\cellcolor[rgb]{0.776,0.922,0.757}}0.556 & {\cellcolor[rgb]{0.776,0.922,0.757}}0.618 & {\cellcolor[rgb]{0.949,0.957,0.969}}\textbf{-} & {\cellcolor[rgb]{0.776,0.922,0.757}}0.601 & {\cellcolor[rgb]{0.776,0.922,0.757}}0.574 & {\cellcolor[rgb]{0.949,0.957,0.969}}\textbf{-} & {\cellcolor[rgb]{0.882,0.957,0.906}}0.541 & {\cellcolor[rgb]{0.776,0.922,0.757}}0.627 & {\cellcolor[rgb]{0.949,0.957,0.969}}\textbf{-} & {\cellcolor[rgb]{0.776,0.922,0.757}}0.611 \\
 & \textbf{\textbf{Hijacking UB}} & {\cellcolor[rgb]{0.627,0.867,0.592}}0.875 & {\cellcolor[rgb]{0.949,0.957,0.969}}\textbf{\textbf{-}} & {\cellcolor[rgb]{0.627,0.867,0.592}}0.874 & {\cellcolor[rgb]{0.627,0.867,0.592}}0.881 & {\cellcolor[rgb]{0.949,0.957,0.969}}\textbf{\textbf{-~}} & {\cellcolor[rgb]{0.627,0.867,0.592}}0.884 & {\cellcolor[rgb]{0.627,0.867,0.592}}0.871 & {\cellcolor[rgb]{0.949,0.957,0.969}}\textbf{\textbf{-~}} & {\cellcolor[rgb]{0.627,0.867,0.592}}0.860 & {\cellcolor[rgb]{0.627,0.867,0.592}}0.852 & {\cellcolor[rgb]{0.949,0.957,0.969}}\textbf{\textbf{-~}} & {\cellcolor[rgb]{0.627,0.867,0.592}}0.843 \\ 
\hline
\multirow{4}{*}{\begin{tabular}[c]{@{}c@{}}\textbf{Hijacking Ethnicity~}\\\textbf{Accuracy}\end{tabular}} & \textbf{\textbf{\textbf{\textbf{Hijacking LB}}}} & \multicolumn{3}{c|}{{\cellcolor[rgb]{0.945,0.98,0.941}}0.200} & \multicolumn{3}{c|}{{\cellcolor[rgb]{0.945,0.98,0.941}}0.200} & \multicolumn{3}{c|}{{\cellcolor[rgb]{0.945,0.98,0.941}}0.200} & \multicolumn{3}{c}{{\cellcolor[rgb]{0.945,0.98,0.941}}0.200} \\
 & \textbf{\textbf{\textbf{\textbf{SnatchML}}}\textbf{~(}Logits)} & {\cellcolor[rgb]{0.882,0.957,0.906}}0.384 & {\cellcolor[rgb]{0.882,0.957,0.906}}0.274 & {\cellcolor[rgb]{0.949,0.957,0.969}}\textbf{-} & {\cellcolor[rgb]{0.882,0.957,0.906}}0.323 & {\cellcolor[rgb]{0.882,0.957,0.906}}0.275 & {\cellcolor[rgb]{0.949,0.957,0.969}}\textbf{-} & {\cellcolor[rgb]{0.776,0.922,0.757}}0.360 & {\cellcolor[rgb]{0.882,0.957,0.906}}0.266 & {\cellcolor[rgb]{0.627,0.867,0.592}}\textbf{-} & {\cellcolor[rgb]{0.808,0.949,0.843}}0.351 & {\cellcolor[rgb]{0.808,0.949,0.843}}0.303 & {\cellcolor[rgb]{0.949,0.957,0.969}}\textbf{-} \\
 & \textbf{\textbf{\textbf{\textbf{SnatchML}}}\textbf{~(}FV)} & {\cellcolor[rgb]{0.776,0.922,0.757}}0.397 & {\cellcolor[rgb]{0.776,0.922,0.757}}0.310 & {\cellcolor[rgb]{0.949,0.957,0.969}}\textbf{-} & {\cellcolor[rgb]{0.776,0.922,0.757}}0.401 & {\cellcolor[rgb]{0.776,0.922,0.757}}0.409 & {\cellcolor[rgb]{0.949,0.957,0.969}}\textbf{-} & {\cellcolor[rgb]{0.882,0.957,0.906}}0.346 & {\cellcolor[rgb]{0.776,0.922,0.757}}0.316 & {\cellcolor[rgb]{0.949,0.957,0.969}}\textbf{-} & {\cellcolor[rgb]{0.776,0.922,0.757}}0.415 & {\cellcolor[rgb]{0.776,0.922,0.757}}0.326 & {\cellcolor[rgb]{0.949,0.957,0.969}}\textbf{-} \\
 & \textbf{\textbf{Hijacking UB}} & {\cellcolor[rgb]{0.627,0.867,0.592}}0.754 & {\cellcolor[rgb]{0.627,0.867,0.592}}0.766 & {\cellcolor[rgb]{0.949,0.957,0.969}}-~ & {\cellcolor[rgb]{0.627,0.867,0.592}}0.746 & {\cellcolor[rgb]{0.627,0.867,0.592}}0.759 & {\cellcolor[rgb]{0.949,0.957,0.969}}-~ & {\cellcolor[rgb]{0.627,0.867,0.592}}0.755 & {\cellcolor[rgb]{0.627,0.867,0.592}}0.758 & {\cellcolor[rgb]{0.949,0.957,0.969}}-~ & {\cellcolor[rgb]{0.627,0.867,0.592}}0.698 & {\cellcolor[rgb]{0.627,0.867,0.592}}0.710 & {\cellcolor[rgb]{0.949,0.957,0.969}}-~ \\
\hline
\end{tabular}

}
\end{table*}

%\subsection{Context}
In this section, we consider an application that predicts personal attributes (e.g., age, gender, ethnicity) from facial images. An example of this application is Microsoft Azure's Face \cite{AzureFace} as an MLaaS API, where users can query one specific model (e.g., for age estimation) using facial images. Given an original task $\mathcal{T}$ (age estimation), an adversary aims to hijack the model to infer other personal attributes like gender or ethnicity. The implications of this hijacking attack can be critical, mainly if users don't consent to using their personal data.  Unauthorized inference of gender and ethnicity from this data could constitute a significant privacy violation, potentially leading to discriminatory practices, especially against individuals from marginalized ethnic or gender groups.

%, who may face unfair treatment and skewed forecasts. 
%In addition, imprecise or prejudiced forecasts resulting from the appropriate model might harm the standing of the company or platform using it, weakening user confidence and eventually reducing engagement and utilization. 

\subsection{Setup}
 
\noindent\textbf{UTKface Dataset \cite{zhang2017age}:} a collection of $20,000$ facial images of individuals ranging in age from $1$ to $116$ years old. Each image is labeled with information on age (6 classes), gender (2 classes), and ethnicity (5 classes).
% as follows: \textcolor{red}{HB: Can we just mention number of classes for each?}\\
% \noindent\textbf{Age}: '\texttt{Newborns/Toddlers}', '\texttt{Pre-adolescence}', '\texttt{Teenagers}', '\texttt{Young Adults}', '\texttt{Middle-aged}', and '\texttt{Seniors}'.\\
% \noindent\textbf{Gender}: '\texttt{Male}' and '\texttt{Female}'.\\
% \noindent\textbf{Ethnicity}: '\texttt{White}', '\texttt{Black}', '\texttt{Asian}', '\texttt{Indian}', and '\texttt{Other}'. 
% (e.g., Hispanic, Latino, Middle Eastern).
%\hl{@Mahmoud : do the ethnicity classes here named by the dataset? (because indian is asian)}
 We consider the same model architectures as in Section \ref{subsec:setup1} and %, trained using a learning rate of $10^{-3}$ and Adam optimizer. % for $10$ epochs for 2D-CNN, $15$ epochs for ResNet-9 and MobileNet, and 30 epochs for Transformer. %The dataset was divided into $70\%$ for training, $10\%$ for validation and $20\%$ for testing. 
 compare  our results with the hijacking upper bound (UB), which corresponds to an unconstrained version of \cite{salem2021get} without covertness requirement. %, which is practically equivalent to freely poisoning the model by training it on both the original and hijacking tasks. %to hijack it at in
\subsection{Cross-attribute hijacking: age, gender and ethnicity}
The adversary's goal is to hijack a model -- exclusively trained on one of the three tasks-- to infer other personal attributes for which the victim model has not been trained. Specifically, from an unknown facial image query submitted by an unknown user to an age estimation model and by only using the original model's BEK, the adversary aims to infer the gender and ethnicity of the user. For a comprehensive study, we test all the possible combinations of (age, gender, and ethnicity) as original and hijacking tasks. %Following the \textit{leave-one-out} on the possible combinations, we exclusively train a model on one task and test \ourapproach~on the two remaining tasks. 
% Following our threat model, we conduct the hijacking attacks in ($\emph{i}$) \textit{black-box} setting -- using the output Logits) and ($\emph{ii}$) \textit{white-box} setting -- using the feature vector (FV) from the last layer of the original model). We also employ a similarity-based distance metric (cosine) for the pairwise comparison between the FV/Logits from unknown query facial images and the FV/Logits obtained by querying the original model with in-the-wild facial images from the hijacking database. We use the facial images from the test dataset as targets for the hijacking attack.
% Following our threat model, we run \ourapproach~using the facial images from the test dataset.
 
Table \ref{tab:utk_attack} provides an overview on the experimental results with random guess probability as a lower bound. 
 Different tasks have varying levels of difficulty that can be related to their inherent complexity. For example, age recognition is a challenging task; a 17-year-old 'teenager' and an 18-year-old  'young adult' are not easily distinguished.
%Age categorization could have been divided into three classes following prior works (\cite{jung2021fair}, \cite{cha2024learning}), but we opted for a higher challenge. For instance, following the three-class division, we achieved an accuracy of 73.5`\% for the white-box attack on the 2D-CNN model trained for the gender prediction task. 
The fact that models trained for age prediction as the original task achieve accuracy ranging from $63\%$ (Transformer) to $70\% $ (ResNet-9) illustrates the task complexity and highlights the relative success of the attack.
Similar conclusions can be drawn regarding hijacking the model for recognizing ethnicity. \ourapproach~achieves over $40\%$ accuracy under the white-box setting for the Transformer and ResNet-9 models for a hijacking UB of $75\%$.
However, we observe that hijacking 'Gender' is more challenging, and the attack has shown its limitations despite its accuracy consistently exceeding random guessing probability. %\ihsen{\textbf{@me todo: comment on UB--}} 
%We expectedly note that the white-box setting consistently delivers higher performance. For instance, ResNet-9, trained for gender recognition, can be hijacked to predict age and ethnicity with an accuracy of  $\sim 45\%$, while the random guessing probability for this task is $16.60\%$ and the UB hijacking accuracy is $66.89\%$.

\section{Scenario 3: Pneumonia Diagnosis}\label{sec:scenario3_Pneu}
%\section{Disease detection to disease classification}

%\subsection{Context}
This section illustrates our attack with a use-case in the healthcare domain, where the original task is Pulmonary Disease Diagnosis (PDD) using chest X-ray images, i.e., a binary classification of whether the patient has 'Pneumonia'. Such a model can be deployed within MLaaS API such as \textit{Google Healthcare API} \cite{GoogleHealth}. Patients or practitioners can query the model with X-ray images and receive feedback on potential pneumonia diagnoses. Our attack investigates the possibility of inferring more information on the type of disease using the same model. Specifically, the attacker tries to infer a hijacking task on samples identified with pneumonia to infer  whether the individual's infection is from a viral or bacterial origin.  We assume that the adversary has access to a database of X-ray images labeled with the type of infection. These images, which may be publicly available, are not necessarily associated with the individuals querying the PDD model. 

\subsection{Setup}

\noindent\textbf{Chest X-Ray Images (Pneumonia) Dataset.} This dataset \cite{kermany2018identifying} consists of $5,863$ X-ray images categorized into '\texttt{Pneumonia}' and '\texttt{Normal}' classes. Two expert physicians graded the diagnoses, and a third expert validated the evaluation set to ensure diagnostic accuracy. The '\texttt{Pneumonia}' category is further labeled into '\texttt{Viral}' and '\texttt{Bacterial}' subcategories. 
% This dataset is organized into training, testing, and validation sets, providing a comprehensive resource for training and evaluating AI systems in pediatric chest X-ray analysis.
We consider the same back-bone architectures as in Section \ref{subsec:setup1} with slight modifications to enable the (original) binary classification. %We train the models using a learning rate of $0.001$ and Adam as optimizer. % for $50$ epochs with early stopping. %We use the same data split for training, validation, and test, as provided by the original Chest X-Ray Images dataset \cite{kermany2018identifying}.

\subsection{PDD Models recognize more than Pneumonia}
\begin{figure}[t]
  \centering
  \includegraphics[width=\linewidth]{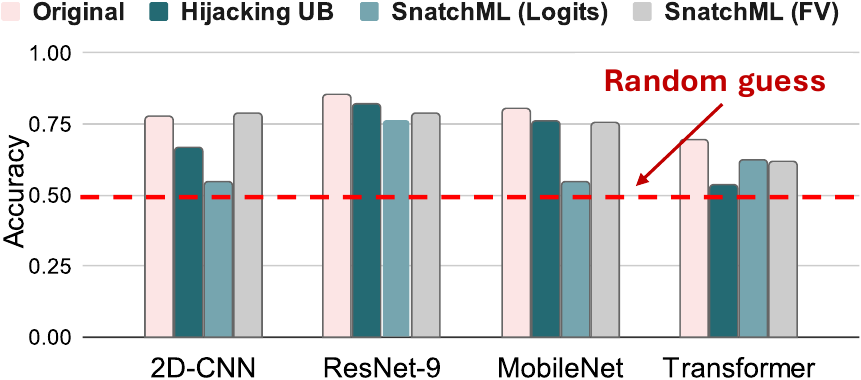}
    \caption{PDD Attack: Accuracy of the original task and hijacking attacks. Random guess probability is at 0.5.
    %\ihsen{please change the reference to 'Hijacking UB'--}. --> Done
    }
  \label{fig:pathology_acc}
\end{figure}

%\textcolor{blue}{\textit{Ihsen--} \hl{Here comes the experimental results and analysis}}
The adversary aims to hijack a PDD model to extract more information on the victim's health record. Specifically, from unknown query X-ray samples, only using the PDD model's output, the attacker aims to infer the pulmonary infection type (i.e., viral or bacterial). %In 
Figure~\ref{fig:pathology_acc} shows the accuracy results of \ourapproach~ under different settings. We observe that the PDD model achieves an accuracy up to $\sim 85\%$ on the original task. Interestingly, the accuracy of the hijacking task (i.e., type of pneumonia infection) is comparable to the accuracy of the model in the original task. Among all architectures, \ourapproach~was particularly successful on ResNet-9 under both white and black-box settings, reaching an attack accuracy of $76\%$ and $78\%$, respectively. Despite its inherited problem of learning complex tasks on small datasets, the Transformer achieves a hijacking accuracy of $62.5\%$. The disparity in attack efficiency as a function of the model architecture for the same settings provides an interesting insight on the importance of the model hyperparameters in the underlying phenomena leading to the hijacking success. Note that under white-box and for all models, there's a strong correlation between the accuracy of the original and hijacking tasks. This further emphasizes the overlap in the learning dynamics for the original and hijacking tasks. \ourapproach~achieves results that are close to the hijacking UB despite being with no training access. %We noticed that the hijacking UB was unsuccessful for the transformer in this setting

\section{Scenario 4: ECG Diagnosis}\label{sec:scenario4_ECG}
%\subsection{Context}
In this section, we evaluate our attack on a modality other than images. Specifically, we focus on ECG signals, which capture cardiac electrical activity over time and produce one-dimensional signal data. To assess the susceptibility of ECG signals to hijacking attacks at inference time, we consider the following settings:

%by examining models trained to classify ECG signals into various categories and assessing their vulnerability to hijacking during inference.
%In this section, we focus on an application in the medical sector, specifically the electrocardiogram (ECG). An ECG is a test that records the electrical activity of the heart over a period of time, producing one-dimensional signal data. So this scenario is different compared to previous ones since we are leveraging signals rather than images. This scenario involves examining models trained to classify ECG signals into various categories and assessing their vulnerability to hijacking during inference.

\noindent \textbf{\underline{Case (\emph{i})}}: The original task $\mathcal{T}$ is ECG-based arrhythmia classification into $5$ categories: 'normal' and $4$ anomalous categories. The adversary’s goal is to hijack a model pre-trained for arrhythmia detection to perform \textit{biometric identification}, attempting to identify individuals from unknown ECG queries. These ECG signals are sourced from a dataset different from the one used for the original task.

    %Secondly, we explore the model's capability to classify speech sequences (recordings of words), extending its functionality beyond its original intent. 
    %this example represents a case of a cross-modality attack, where 
    %This means that the task performed during the attack spans multiple modalities or types of data. 
    %the original model ECG signals classification is hijacked to classify speech sequences (audio signals). 
    %This extension of functionality expands the model's capabilities beyond its initial scope, highlighting its susceptibility to misuse in unintended tasks.

\noindent \textbf{\underline{Case (\emph{ii})}}: The original task is a binary classification of normal versus arrhythmic signals. The adversary aims to repurpose the model to identify specific types of cardiac irregularities within the ECG data, categorizing them into $5$ predefined classes. 
%This involves leveraging the model beyond its original scope to identify and categorize specific types of cardiac irregularities within ECG data, thereby assessing its adaptability and resilience to being repurposed.

\subsection{Setup}
\noindent\textbf{MIT-BIH Arrhythmia Dataset \cite{moody2001impact}:} includes ECG recordings from $47$ participants. % across 48 half-hour sessions with dual channels. Each channel was recorded at 360 samples per second with 11-bit resolution and a 10 mV range. Annotations, totaling approximately 
The dataset contains over $110,000$ samples, categorized into $5$ classes: N (Normal), S (Supraventricular ectopic beat), V (Ventricular ectopic beat), F (Fusion beat), and Q (Unknown beat).

%\vspace{0.2cm}

\noindent\textbf{ECG-ID \cite{lugovaya2005biometric}:} comprises $310$ ECG recordings from $90$ individuals and labeled by identity. %Each recording captures ECG lead I data for 20 seconds at 500 Hz. Annotations include R- and T-wave peaks, with metadata such as age, gender, and recording dates for each identity.
% \begin{figure}[ht]
% \centering
% \includegraphics[width=0.9\linewidth]{attack_figures/voice_signal.png}
% \caption{Visualization of a voice signal sample From FSDD dataset. \textcolor{blue}{Combine this with Figure 9 vertically.}}
% \label{fig:voice_signal}
% \end{figure}
\noindent To process ECG 1D-inputs, we use $3$ models: 1D-CNN, 1D-ResNet-18, and 1D-Inception.

\subsection{Hijacking arrhythmia classifiers for biometric identification}
\begin{figure}[!tp]
  \centering
  \includegraphics[width=1.\linewidth]{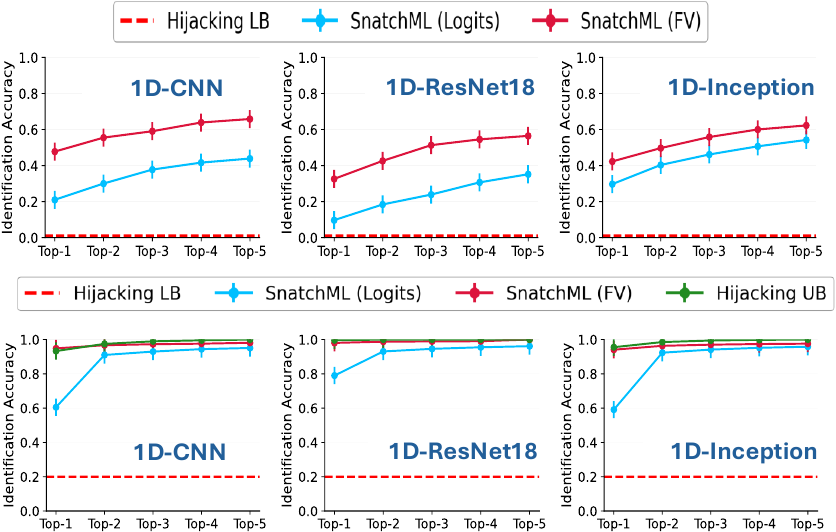}
  \caption{SnatchML performance on hijacking ECG: \textit{1st Row}: biometric identification on ECG-ID dataset. 
  \textit{2nd Row}: Binary classification of cardiac abnormalities on the MIT-BIH Arrhythmia dataset.
 }
  \label{fig:ecg_results}
\end{figure}

The results of the first scenario are depicted in the first row of Figure \ref{fig:ecg_results} across the different architectures (1D-CNN, 1D-Inception, and 1D-ResNet-18). For example, 1D-Inception ranges from $29.68\%$ (Top-1) to $54.19\%$ (Top-5) when the attack is black-box. These results illustrate the models' effectiveness in biometric identification compared to a random accuracy of $0.11\%$. This underscores the models' capability to generalize beyond their original task of ECG classification, posing notable implications for security and privacy. % in sensitive biometric applications. 

\subsection{Binary Classification Models learn arrhythmia categories}
%To evaluate the second scenario, we follow the same approach in sections (\ref{sec:scenario1ER} and \ref{sec:scenario1ER}). 
Figure \ref{fig:ecg_results} demonstrates the vulnerability of the three models in a hijacking scenario with binary classification  of \texttt{'normal'} versus \texttt{'anomalous'} ECG as original task. \ourapproach~ aims to hijack the models for ECG classification into five distinct categories of Arrhythmia (see the second row of Figure  \ref{fig:ecg_results}). The models achieve relatively high accuracy reaching $79\%$ 1D-ResNet-18, $59\%$ for 1D-Inception, and $60\%$ for 1D-CNN  while the hijacking UB accuracy is up to $98\%$ . This setting underscores the models' ability to generalize across different types of cardiac irregularities, highlighting the hijacking risk in sensitive domains such as healthcare.

% \begin{table}[ht]
% \centering
% \caption{SnatchML performance on hijacking a binary classification ECG model to for cardia abnormalities classification on the MIT-BIH Arrhythmia dataset.}
% \label{fig:ecg_multiclass}
% \scalebox{1.0}{
% \begin{tabular}{llllll} 
% \hline
% \textbf{Hijacking Method} & \textbf{TOP-1} & \textbf{TOP-2} & \textbf{TOP-3} & \textbf{TOP-4} & \textbf{TOP-5} \\ 
% \hline
% Random Guess & \multicolumn{5}{c}{{\cellcolor[rgb]{0.945,0.98,0.941}}0.2000} \\ 
% \hline
% SnatchML, 1D-CNN & {\cellcolor[rgb]{0.808,0.949,0.843}}0.6051 & {\cellcolor[rgb]{0.808,0.949,0.843}}0.9101 & {\cellcolor[rgb]{0.808,0.949,0.843}}0.9298 & {\cellcolor[rgb]{0.808,0.949,0.843}}0.9441 & {\cellcolor[rgb]{0.808,0.949,0.843}}0.9508 \\
% SnatchML, 1D-ResNet-18 & {\cellcolor[rgb]{0.808,0.949,0.843}}0.7899 & {\cellcolor[rgb]{0.808,0.949,0.843}}0.9306 & {\cellcolor[rgb]{0.808,0.949,0.843}}0.9458 & {\cellcolor[rgb]{0.808,0.949,0.843}}0.955 & {\cellcolor[rgb]{0.808,0.949,0.843}}0.9614 \\
% SnatchML, 1D-Inception & {\cellcolor[rgb]{0.808,0.949,0.843}}0.5911 & {\cellcolor[rgb]{0.808,0.949,0.843}}0.9234 & {\cellcolor[rgb]{0.808,0.949,0.843}}0.9414 & {\cellcolor[rgb]{0.808,0.949,0.843}}0.9519 & {\cellcolor[rgb]{0.808,0.949,0.843}}0.9583 \\
% \hline
% \end{tabular}
% }
% \end{table}

% \begin{figure}[ht]
% \centering
% \includegraphics[width=0.9\linewidth]{attack_figures/ecg_2_to_5.png}
% \caption{Evaluation of SnatchML in Extending Binary ECG Classification to Multiclass Classification from the MIT-BIH Arrhythmia dataset.}
% \label{fig:ecg_5}
% \end{figure}

\section{What if the hijacking and original tasks are unrelated?}\label{sec:unrelated}

Previous experiments demonstrate that a model trained for an original task can be effectively hijacked to perform a different task. This finding has so far involved scenarios where the hijacking tasks share semantic relatedness or an overlap in data distribution with the original task. While this represents a significant security threat, the requirement for task relatedness can be viewed as a limitation from an attack perspective. In fact, state-of-the-art model hijacking attacks are more general and are not particularly constrained by the relatedness to the original task.
In this section, we extend our investigation to examine the generalizability of \ourapproach~to hijacking tasks that are unrelated to the original task.

% \noindent \textbf{(ii)} In the second, we consider randomly initialized models, i.e., the parameters are sampled from a random distribution, and run the hijacking attack. \textcolor{blue}{$\leftarrow$} \ihsen{We will not consider this second case here but rather in Section 9.2}
%Previous experiments demonstrate that a model trained for a specific task can be effectively hijacked to perform a different task. This finding is particularly relevant in scenarios where the tasks share semantic relatedness or an overlap in data distribution. Although this poses a significant threat, the requirement for task relatedness can be viewed as a limitation from an attack perspective. In fact, state-of-the-art model hijacking attacks are more versatile and are not limited by relatedness to the original task.

% In this section, we extend our investigation to examine the generalizability of our study to hijacking tasks that are unrelated to the original task

% Surprisingly, we found that the model hijacking is successful even in these settings. %We provide insights on the possible reason behind these findings from the random projection theory. 

\begin{figure}[t]
\centering
\includegraphics[width=\linewidth]{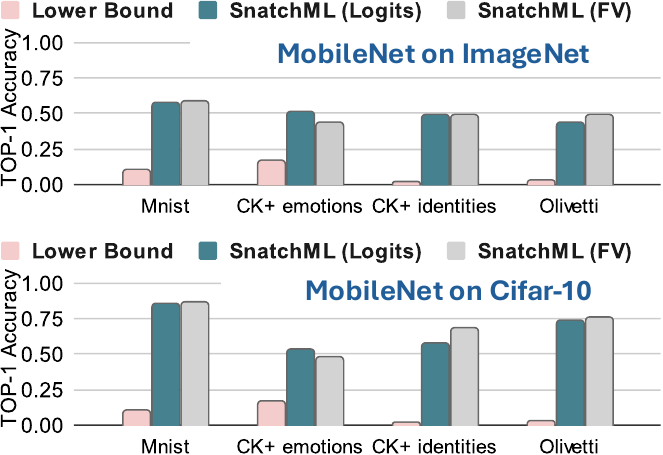}
\caption{Results of \ourapproach~on MobileNet pretrained on ImageNet and Cifar10. The models are hijacked to infer other unrelated tasks defined by their datasets on the x-axis. }%\textcolor{red}{HB: move to appendix?}}
\label{fig:unrelated_tasks}
\end{figure}

\subsection{Unrelated Image Classification Tasks}
\noindent \textbf{Setup.} We evaluate several image classification benchmarks as original tasks. We use a MobileNet pretrained on ImageNet  and on CIFAR-10, from the Pytorch model Zoo \cite{zoo}. These pretrained models are the targets of \ourapproach, which aims to hijack them for unrelated tasks involving different data distributions. Specifically, we hijack the models for: MNIST, CK+ for ER, CK+ for biometric identification, Olivetti for face recognition. We run \ourapproach~on the testset distribution to evaluate the hijacking task performance.

For each hijacking attack, Figure \ref{fig:unrelated_tasks} presents the performance under both black-box and white-box settings, along with the probability of a random guess. The results illustrate the capacity of overparametrised models to extract useful knowledge from a data distribution that is unrelated to their original training dataset. In fact, the \ourapproach~hijacking accuracy is comparable to the state-of-the-art accuracy of tasks like MNIST, when the victim model is trained on ImageNet. %In the following section, we attempt to provide possible explanations for these results.   

\subsection{Cross modality: hijacking ECG for speech recognition}

In the following, we  investigate unrelated tasks where the original task involves a different modality than the hijacking task, i.e., hijacking ECG-based arrhythmia classification for speech recognition.

\noindent \textbf{Setup. }  We evaluate three models pretrained on the MIT-BIH Arrhythmia Dataset: 1D-CNN, ResNet-18, and 1D-Inception.  The adversary wants to hijack the models for speech recognition using the Free Spoken Digit (FSDD) Dataset \cite{fsdd}. The FSDD dataset consists of approximately $2,000$ audio recordings of spoken digits (0-9) by multiple speakers.

\begin{figure}[htp]
  \centering
  \includegraphics[width=1.\linewidth]{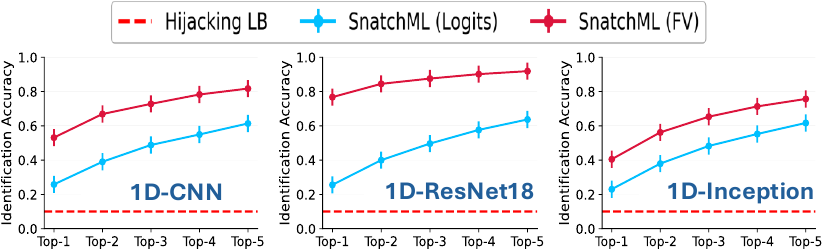}
  \caption{SnatchML performance on hijacking an ECG models for speech classification on Free Spoken Digit dataset. 
 }
  \label{fig:ecg_speech}
\end{figure}

\noindent \textbf{Results. } The experimental results shown in Figure \ref{fig:ecg_speech} highlight the effectiveness of \ourapproach~ in a cross-modality hijacking. These models achieved Top-1 accuracy around $23\%$ to $26\%$ and Top-5 accuracy ranging from $61\%$ to $64\%$.

%By leveraging BEK (logits) and distance metrics like Cosine Similarity, the study demonstrates the unintended capability of these models to generalize across modalities.

% \begin{table}[ht]
% \centering
% \caption{SnatchML performance on hijacking an ECG model for speech classification on Free Spoken Digit dataset.}
% \label{fig:ecg_voice}
% \scalebox{1.0}{
% \begin{tabular}{llllll} 
% \hline
% \textbf{Hijacking Method} & \textbf{TOP-1} & \textbf{TOP-2} & \textbf{TOP-3} & \textbf{TOP-4} & \textbf{TOP-5} \\ 
% \hline
% Random Guess & \multicolumn{5}{c}{{\cellcolor[rgb]{0.945,0.98,0.941}}0.1000} \\ 
% \hline
% SnatchML, 1D-CNN & {\cellcolor[rgb]{0.808,0.949,0.843}}0.2580 & {\cellcolor[rgb]{0.808,0.949,0.843}}0.3901 & {\cellcolor[rgb]{0.808,0.949,0.843}}0.4881 & {\cellcolor[rgb]{0.808,0.949,0.843}}0.5494 & {\cellcolor[rgb]{0.808,0.949,0.843}}0.6136 \\
% SnatchML, 1D-ResNet-18 & {\cellcolor[rgb]{0.808,0.949,0.843}}0.2556 & {\cellcolor[rgb]{0.808,0.949,0.843}}0.3996 & {\cellcolor[rgb]{0.808,0.949,0.843}}0.4963 & {\cellcolor[rgb]{0.808,0.949,0.843}}0.5761 & {\cellcolor[rgb]{0.808,0.949,0.843}}0.6370 \\
% SnatchML, 1D-Inception & {\cellcolor[rgb]{0.808,0.949,0.843}}0.2300 & {\cellcolor[rgb]{0.808,0.949,0.843}}0.3794 & {\cellcolor[rgb]{0.808,0.949,0.843}}0.4823 & {\cellcolor[rgb]{0.808,0.949,0.843}}0.5523 & {\cellcolor[rgb]{0.808,0.949,0.843}}0.6165 \\
% \hline
% \end{tabular}
% }
% \end{table}

\subsection{Features visualization}

\begin{figure}[htp]
  \centering
  \includegraphics[width=0.9\linewidth]{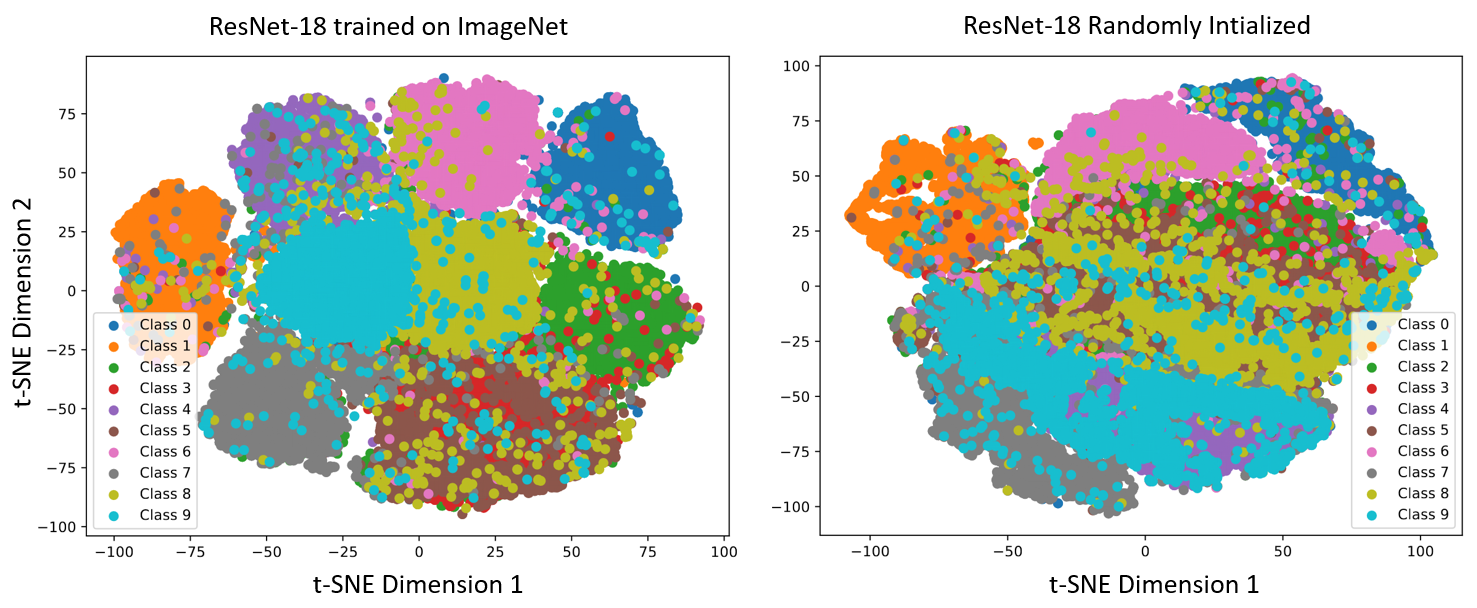}
  \caption{Comparison of t-SNE distributions for MNIST digit classes using ResNet18 pretrained on ImageNet (left) and randomly initialized ResNet18 (right).}
  \label{fig:mnist-tsne}
\end{figure}

To visualize the model's capacity to distinguish classes from data distributions unrelated to the training set, in Figure \ref{fig:mnist-tsne}, we present t-SNE plots comparing the feature distributions of MNIST digit classes using feature vectors extracted from a ResNet-18 model that is pretrained on ImageNet, and a randomly initialized ResNet-18.  We observe distinct clusters corresponding to different digit classes in both cases, despite neither model being trained on the MNIST dataset. This observation confirms the previous results and illustrates the potential for model hijacking, even when the hijacking task is entirely unrelated to the original task. 

\section{What is the impact of task complexity?}

\begin{figure}[tp]
  \centering
  \includegraphics[width=\linewidth]{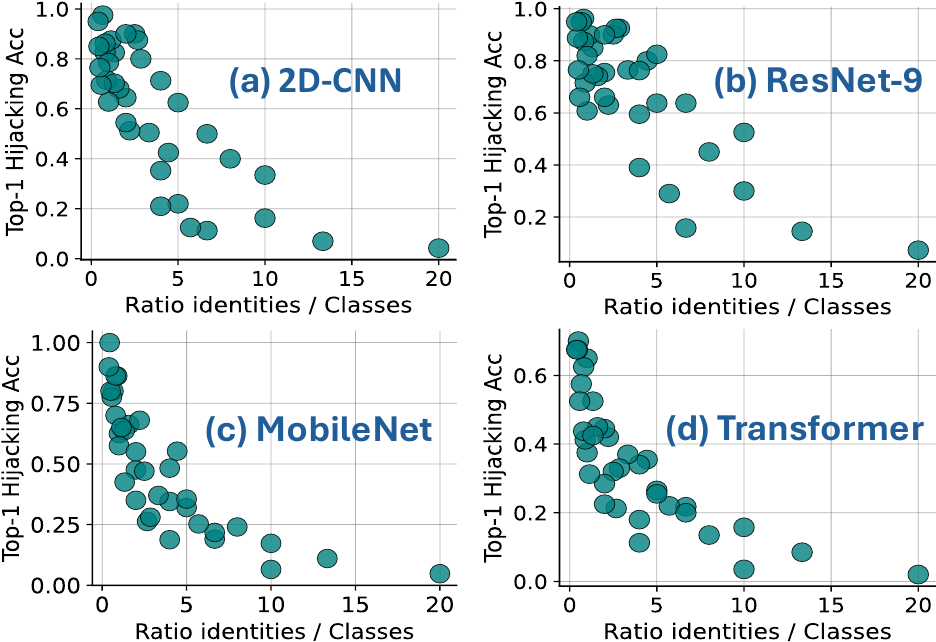}
  \caption{ Effect of complexity ratio $r=m/n$ on the hijacking attack Accuracy. }
  \label{fig:ratio}
\end{figure}
% \noindent \textbf{}
Unlike existing work, our approach enables repurposing a model trained for a $n$-class classification task $\mathcal{T}$ for an $m$-class hijacking task $\mathcal{T'}$, where $n \leq m$. In this section, we propose to investigate the extent of this property. Specifically, we explore the attack performance while varying the relative hijacking task complexity. For a given architecture, we use the number of classes as a proxy of the task complexity and define the complexity ratio $r=m/n$ as a metric to evaluate how difficult is the hijacking attack.  
%From lower to higher hijacking difficulty, we varied both the number of original classes $n$ at training and the number of attack classes $m$. 
The study was conducted using MobileNet, ResNet-9, 2D-CNN, and Transformer models, which were trained on MNIST. The number of training classes $n$ varied from only two digits to models trained to classify all $10$ digits in the complete dataset. Subsequently, the attack was performed to achieve facial recognition using Olivetti dataset, which contains $40$ identities. We conduct the attack with four different $m$ values by sampling different numbers of identities $m=40$, $m=20$, $m=8$, and $m=4$ identities. This resulted in 36 possible combinations, as shown in Figure \ref{fig:ratio}.
The scatter trend of Figure \ref{fig:ratio} reveals a clear inverse relationship between the ratio $m/n$ and the hijacking accuracy. As the ratio increases, the hijacking accuracy decreases. The number of classes in the hijacking task $m$ becomes significantly larger than the number of classes the model was originally trained on $n$, \ourapproach's ability to accurately hijack and classify the new classes diminishes. However, it is interesting to note that depending on the classes combination, \ourapproach~ achieved over $80\%$ Top-1 accuracy of the hijacking task, with $m = 5 \times n$ for ResNet-9, while related work do not work for $r>1$.

%This decline in performance is evident from the descending pattern of the data points, which move from higher accuracy levels on the left (lower $Na/Nt$ ratios) to lower accuracy levels on the right (higher $Na/Nt$ ratios). This behavior highlights the challenges and limitations faced when attempting to repurpose a model trained on fewer classes to handle a more complex classification task with a greater number of classes. Despite this limitation, our method remains the only one, to our knowledge, capable of performing such an attack when $N_a$ is greater than $N_t$. While the hijacking accuracy decreases as the ratio of $N_a$ to $N_t$ increases, the fact that our approach can still manage to hijack the model and perform the classification task, even with more classes than the model was originally trained for, is a significant achievement. 
%Existing hijacking attacks in the literature are constrained by the requirement that $N_t$ must be greater than or equal to $N_a$, making our method especially valuable for applications where the number of target classes exceeds the number of training classes. This capability opens up new possibilities for repurposing trained models in scenarios that were previously considered unfeasible.

\begin{figure*}[ht!]
  \centering
  \includegraphics[width=2.1\columnwidth]{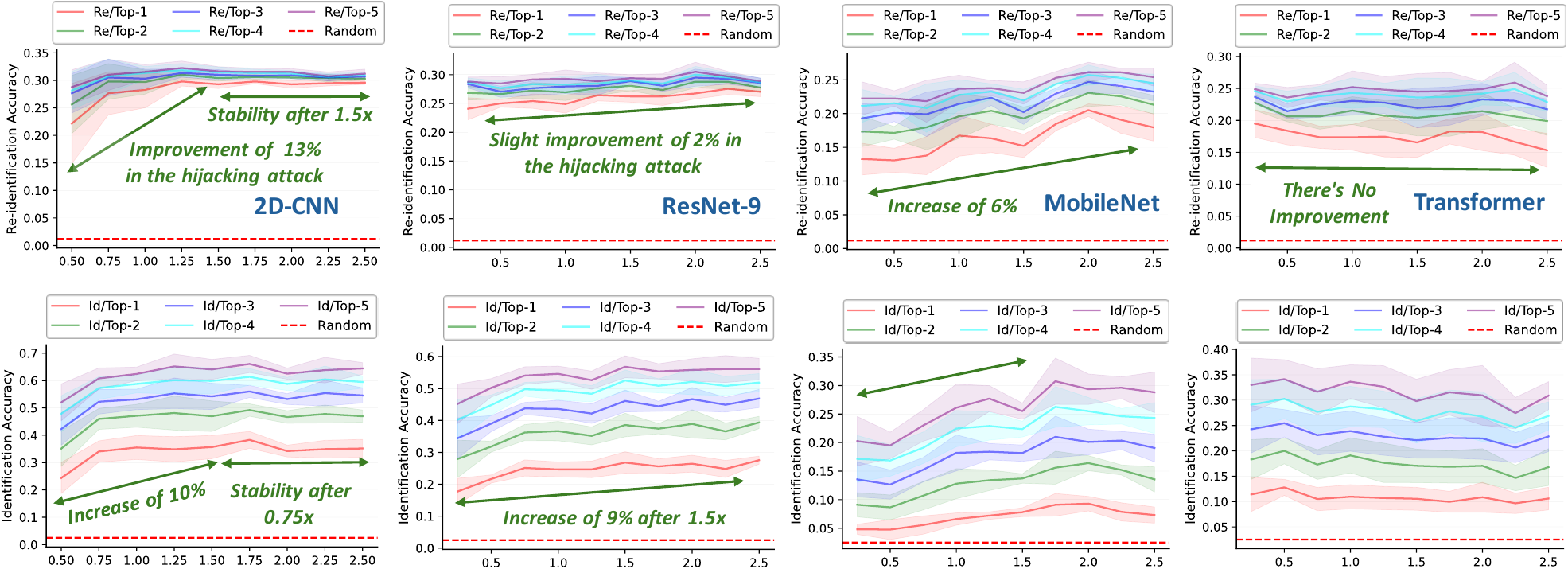}
  \caption{Correlation between the size of benign models trained on original task (ER on CK+ dataset) and their effectiveness in learning hijacking tasks: The results depict the variations in the accuracy of the hijacking tasks for user re-identification (First row) and user identification (Second row) when scaling the ML model size via model's width expansion. The hijacked models are of different architectures (per column): 2D-CNN, ResNet-9, MobileNet, and Transformer. %\halima{Change TR.}
  %\ihsen{\hl{@Halima-- could you please change the transformer to 'no improvement' -- and check Resnet-9, it seems there's a minor improvement. Also in 2D-CNN there is one with min $15\%$ and max $30\%$-- please check the others as well}} \halima{I updated the figure, will update the discussion as well.}
  } 
  \label{fig:overparam_study}
\end{figure*}

\section{Why do models learn more than they should?}\label{sec:explain}

In the previous sections, we show that \ourapproach~can hijack ML models to perform tasks that can be related or entirely unrelated to the original task. While the former is intuitive, the latter is more unexpected. To explore the underlying reasons, we examine two hypotheses: (\emph{i}) Overparametrization of ML model and (\emph{ii}) Random Projection.

\subsection{ Correlation with model size}
\label{subsec:overparams}

ML models may learn features that are not particularly related to the original task. For example, in the context of ER, these models might capture facial characteristics that could inadvertently be used for biometric identification. From a conceptual perspective, such phenomenon might lead to critical security breaches, mainly for not respecting the fundamental principle of "least privilege" in cybersecurity. Practically, the amount of parameters in a ML model reflects the upper bound of its learning capacity. The overparametrization implies that such models contain more parameters than necessary to accomplish the original task. In both software~\cite{wang2013towards} and hardware~\cite{fern2015hardware} systems, \textbf{undefined behavior and don't-care states} have been shown to be potential sources of vulnerabilities.  In the same vein, we believe that this extra-capacity represents undefined behavior and don't-care states with respect to the model that can be exploited by attackers. %We posit that the success of our hijacking attack is related to the overparametrization of victim models. %, i.e., the extra-capacity is in our case exploited to repurpose the original model to perform the hijacking tasks. 
We suggest that the model's capacity to unintentionally learn unnecessary features positively correlates with the accuracy of the hijacking task. %This is particularly supported by the fact that we're exploiting the same extracted feature from the original model to perform hijacking tasks. We assume that within those features, some strongly correlate with the hijacking task (e.g., ER vs. biometric identification). %Consequently, we argue that the dimensionality of the extracted features may impact the likelihood of leaking unnecessary information about users during inference time that may be further exploited for hijacking attacks under black or white box settings.

To verify this hypothesis, we study the correlation between the ML model size and accuracy of the hijacking tasks for \textit{user re-identification} and \textit{biometric identification}. The results of our study are shown in Figure \ref{fig:overparam_study}. We study the case of four ML models, notably 2D-CNN, ResNet-9, MobileNet, and Transformer. We vary the width expansion ratio (i.e., number of channels), from $0.25\times$ to $2.5\times$ to emulate different amounts of model parameters while maintaining the same baseline model architecture. From the reported results in Figure \ref{fig:overparam_study}, we draw the following insights:

\textbf{First}, we notice a positive correlation between the model's width expansion ratio and the hijacking task accuracy. This can be attributed to the increase in the model's capacity to learn facial features from ER training data that are leveraged for user identification in the hijacking attack. For instance, for a re-identification attack with 2D-CNN, the learning capacity of the hijacking task improves by around $13\%$ under $1.25\times$ width expansion and experiences stability afterward. With MobileNet, we notice a steady increase in the hijacking task accuracy, achieving a peak of $6\%$ improvement. The low hijacking accuracy of MobileNet compared to other models can be explained by the compactness of this model, which translates into a lower capacity to learn extra tasks.

\textbf{Second}, we also observe that different model architectures exhibit different behaviors when scaling the model size. This is particularly clear for ResNet-9 and Transformer models (first row of Figure \ref{fig:overparam_study}). The over-parametrized nature of their architecture can explain this trend, even at low-width expansion ratios. Thus, further up-scaling their number of parameters does not improve the hijacking task.

\textbf{Third}, the observed trends of the model size and the accuracy of the hijacking task differ w.r.t the task complexity. For instance, ResNet-9 shows slight improvement of $2\%$ in the hijacking attack for user re-identification under overparameterized regime. For the identification task in Olivetti, ResNet-9 depicts higher improvement of $9\%$ in the hijacking attack ($2^{nd}$ row, $1^{st}$ and $2^{nd}$ columns in Figure \ref{fig:overparam_study}). The re-identification task (i.e., the ER training dataset) includes $89$ user identities, making the \textit{re-identification} task much more challenging than the \textit{identification} task, which involves $\sim40$ user identities. %, presents a relatively simpler case in which the hijacking task accuracy is higher than in the former case. %Hence, for ER systems, it is recommended that datasets be built with a diverse user base to prevent the model from learning specific characteristics of the users that can be exploited for hijacking purposes.

% \textcolor{blue}{Maybe we can move the preliminary analysis here as explanation instead of an intuition-- }

%Nota: in this particular case, does the training mechanism impact the generalisation and hence the hijacking vulnerability?

\subsection{General case: Universality of Random DNNs}\label{subsec:rnd_proj}

% \textcolor{blue}{random projection theorem}
The general case corresponds to a model that is hijacked for a task that is totally unrelated to the original task. The rationale by which features learnt by the model on the original task overlap with the hijacking task does not hold under these assumptions. In fact, as far as the hijacking task is concerned, this case is similar to a randomly initialised model. Our hypothesis is that the relatively high accuracy on the hijacking task can be explained by the universal approximation capabilities of random DNNs. We recall the Johnson-Lindenstrauss Lemma \cite{JL-lemma} which states that a set of points in a high-dimensional space can be projected into a lower-dimensional space while approximately preserving the pairwise distances between the points. While this lemma might give an interesting insight, it cannot rigorously explain the behavior of random ML models because of their non-linearity. However,  several works such as \cite{Rand_uni, randomNN, liao18, rahimi} investigated the capabilities of random neural networks. Particularly, Giryes et. al. \cite{Rand_uni} prove that DNN preserve the metric structure of the data as it propagates along the layers. Interestingly for our work, they show that networks tend to decrease the Euclidean distances between points with a small angle between them (“same class”), more than the distances between points with large angles between them (“different classes”). Further, the theoretical underpinnings provided by \cite{Rand_uni} demonstrate that deep random networks can act as powerful feature extractors, where the depth and width of the network play crucial roles in enhancing the representational power of the features.

\begin{figure}[tp]
  \centering
  \includegraphics[width=\linewidth]{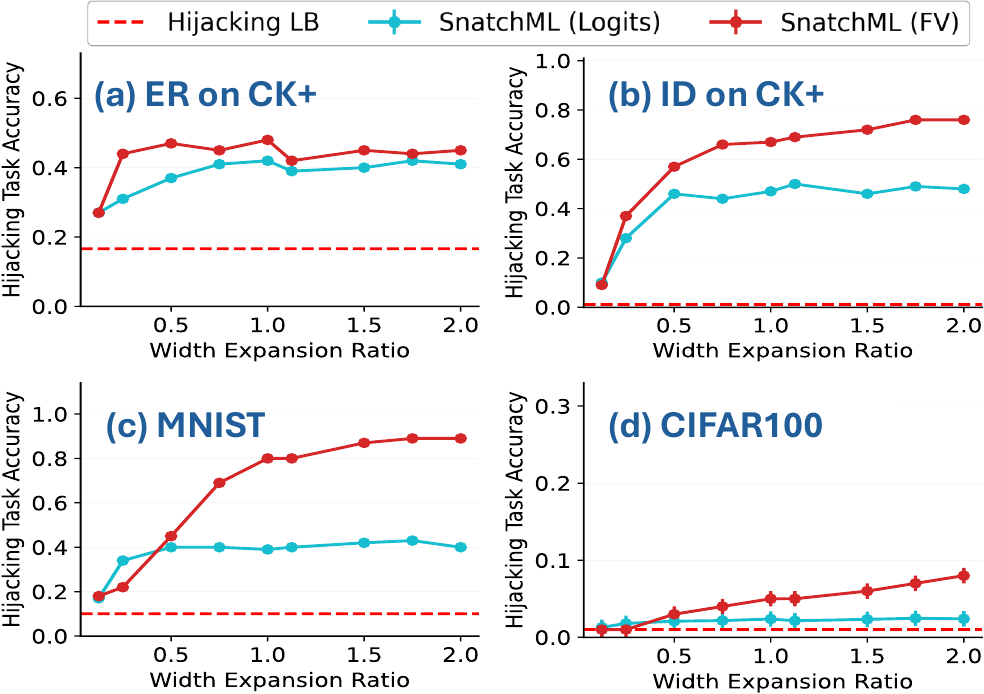}
  \caption{Hijacking attack on a randomly initialized ResNet18 model to perform ER, Identification on CK+ and image classification on MNIST and CIFAR100.}
  \label{fig:random}
\end{figure}

% stated below: 

% \begin{lemma}[Johnson-Lindenstrauss]
% For any $0 < \epsilon < 1$ and any integer $n$, let $k$ be a positive integer such that 
% \[
% k \geq \frac{4 \log n}{\epsilon^2 / 2 - \epsilon^3 / 3}.
% \]
% Then for any set $V$ of $n$ points in Euclidean space $\mathbb{R}^d$, there exists a linear map $f : \mathbb{R}^d \rightarrow \mathbb{R}^k$ such that for all $u, v \in V$,
% \[
% (1-\epsilon) \| u - v \|^2 \leq \| f(u) - f(v) \|^2 \leq (1+\epsilon) \| u - v \|^2.
% \]
% \end{lemma}

% The JL-lemma states that a set of points in a high-dimensional space can be projected into a lower-dimensional space while approximately preserving the pairwise distances between the points. While this lemma cannot rigurously explain the behavior of random ML models because of their non-linearity,  \cite{Rand_uni} built on it to prove that 

While the properties of random neural networks is extensively studied in the ML community, our work shows the potential security threats that can emanate from these properties. To illustrate this perspective, we run our hijacking attack on totally random neural networks, i.e., the parameters are sampled from a random Gaussian distribution. The results shown in Figure \ref{fig:random} are coherent with our observations in Section \ref{sec:unrelated}. We provide a visualization of this setting in the latent space in Figure \ref{fig:mnist-tsne} in the Appendix. 

%preservation of distance is crucial because it maintains the relative geometrical and statistical properties of the data in the latent space, which potentially leads to being efficiently classified based on the distance analysis such as \ourapproach. Therefore, a random neural network can behave as a random projector in the latent space. 

% A similar observation has been formalised in \cite{randomNN} by approximating random deep neural networks with Gaussian Processes. 

%\textcolor{red}{\hl{Here come the results with random neural networks}}

%\input{sec/08_attack_analysis}
\section{Countermeasures}

% We propose $2$ possible countermeasures to mitigate the threat of \ourapproach: 

\begin{table*}[t]
\centering
\caption{Model compression: accuracy of the original and hijacking tasks on different datasets and models.} %For each case, we report accuracy values for the baseline model $f(.)$ and compact model $f_{cmp}(.)$ found by our exhaustive search for width expansion ratios ($m$). We evaluate the hijacking under black-box (Logits) and white-box (FV) settings.}
\label{tab:comp_results}
\scalebox{1}{
\begin{tabular}{cc|ccc|cc|cc} 
\hline
\multirow{2}{*}{\begin{tabular}[c]{@{}c@{}}\textbf{Hijacking Task}\\\textbf{(Dataset)}\end{tabular}} & \multirow{2}{*}{\textbf{ Model}} & \multicolumn{3}{c|}{\textbf{Original Task Accuracy}} & \multicolumn{4}{c}{\textbf{\ourapproach: Hijacking Task Accuracy}} \\ 
\cline{3-9}
 &  & $f(.)$ & $f_{cmp}(.)$ & \multicolumn{1}{l|}{\textbf{Expansion $m$}} & \textbf{Logit of~}$f(.)$ & \textbf{\textbf{Logit of}}$f_{cmp}(.)$ & \textbf{FV of~}$f(.)$ & \textbf{FV of}$f_{cmp}(.)$ \\ 
\hline
\multirow{3}{*}{\begin{tabular}[c]{@{}c@{}}\textbf{Re-identification}\\\textbf{(CK+ )}\end{tabular}} & 2D-CNN & {\cellcolor[rgb]{0.984,0.937,0.937}}0.967 & {\cellcolor[rgb]{0.984,0.937,0.937}}0.941 & 0.5$\times$ & {\cellcolor[rgb]{0.882,0.957,0.906}}0.298 & {\cellcolor[rgb]{0.776,0.922,0.757}}0.244 & {\cellcolor[rgb]{0.882,0.957,0.906}}0.329 & {\cellcolor[rgb]{0.776,0.922,0.757}}0.320 \\
 & Resnet-9 & {\cellcolor[rgb]{0.984,0.937,0.937}}0.937 & {\cellcolor[rgb]{0.984,0.937,0.937}}0.935 & 0.1$\times$ & {\cellcolor[rgb]{0.882,0.957,0.906}}0.257 & {\cellcolor[rgb]{0.776,0.922,0.757}}0.217 & {\cellcolor[rgb]{0.776,0.922,0.757}}0.293 & {\cellcolor[rgb]{0.882,0.957,0.906}}0.294 \\
 & Mobilenet & {\cellcolor[rgb]{0.984,0.937,0.937}}0.939 & {\cellcolor[rgb]{0.976,0.867,0.863}}0.915 & 0.45$\times$ & {\cellcolor[rgb]{0.882,0.957,0.906}}0.148 & {\cellcolor[rgb]{0.776,0.922,0.757}}0.143 & {\cellcolor[rgb]{0.882,0.957,0.906}}0.176 & {\cellcolor[rgb]{0.776,0.922,0.757}}0.163 \\ 
\hline
\multicolumn{1}{l}{} & \multicolumn{1}{l}{} & \multicolumn{1}{l}{} & \multicolumn{1}{l}{} & \multicolumn{1}{l}{} & \multicolumn{1}{l}{} & \multicolumn{1}{l}{} & \multicolumn{1}{l}{} & \multicolumn{1}{l}{} \\ 
\hline
\multirow{3}{*}{\begin{tabular}[c]{@{}c@{}}\textbf{\textbf{Identification}}\\\textbf{(Olivetti~\textbf{})}\end{tabular}} & 2D-CNN & {\cellcolor[rgb]{0.984,0.937,0.937}}0.967 & {\cellcolor[rgb]{0.984,0.937,0.937}}0.941 & 0.5$\times$ & {\cellcolor[rgb]{0.882,0.957,0.906}}0.319 & {\cellcolor[rgb]{0.776,0.922,0.757}}0.289 & {\cellcolor[rgb]{0.882,0.957,0.906}}0.564 & {\cellcolor[rgb]{0.776,0.922,0.757}}0.523 \\
 & Resnet-9 & {\cellcolor[rgb]{0.984,0.937,0.937}}0.937 & {\cellcolor[rgb]{0.984,0.937,0.937}}0.935 & 0.1$\times$ & {\cellcolor[rgb]{0.882,0.957,0.906}}0.232 & {\cellcolor[rgb]{0.776,0.922,0.757}}0.144 & {\cellcolor[rgb]{0.882,0.957,0.906}}0.479 & {\cellcolor[rgb]{0.776,0.922,0.757}}0.359 \\
 & Mobilenet & {\cellcolor[rgb]{0.984,0.937,0.937}}0.939 & {\cellcolor[rgb]{0.976,0.867,0.863}}0.915 & 0.45$\times$ & {\cellcolor[rgb]{0.882,0.957,0.906}}0.074 & {\cellcolor[rgb]{0.776,0.922,0.757}}0.070 & {\cellcolor[rgb]{0.882,0.957,0.906}}0.076 & {\cellcolor[rgb]{0.776,0.922,0.757}}0.066 \\ 
\hline
\multicolumn{1}{l}{} & \multicolumn{1}{l}{} & \multicolumn{1}{l}{} & \multicolumn{1}{l}{} & \multicolumn{1}{l}{} & \multicolumn{1}{l}{} & \multicolumn{1}{l}{} & \multicolumn{1}{l}{} & \multicolumn{1}{l}{} \\ 
\hline
\multirow{3}{*}{\begin{tabular}[c]{@{}c@{}}\textbf{Type of Pneumonia}\\\textbf{(Chest X-ray~)}\end{tabular}} & 2D-CNN & {\cellcolor[rgb]{0.984,0.937,0.937}}0.776 & {\cellcolor[rgb]{0.984,0.937,0.937}}0.764 & 0.1$\times$ & {\cellcolor[rgb]{0.776,0.922,0.757}}0.544 & {\cellcolor[rgb]{0.882,0.957,0.906}}0.564 & {\cellcolor[rgb]{0.882,0.957,0.906}}0.787 & {\cellcolor[rgb]{0.776,0.922,0.757}}0.759 \\
 & Resnet-9 & {\cellcolor[rgb]{0.984,0.937,0.937}}0.853 & {\cellcolor[rgb]{0.984,0.937,0.937}}0.851 & 0.1$\times$ & {\cellcolor[rgb]{0.882,0.957,0.906}}0.764 & {\cellcolor[rgb]{0.776,0.922,0.757}}0.541 & {\cellcolor[rgb]{0.882,0.957,0.906}}0.787 & {\cellcolor[rgb]{0.776,0.922,0.757}}0.756 \\
 & Mobilenet & {\cellcolor[rgb]{0.984,0.937,0.937}}0.801 & {\cellcolor[rgb]{0.976,0.867,0.863}}0.780 & 0.1$\times$ & {\cellcolor[rgb]{0.882,0.957,0.906}}0.544 & {\cellcolor[rgb]{0.776,0.922,0.757}}0.482 & {\cellcolor[rgb]{0.882,0.957,0.906}}0.751 & {\cellcolor[rgb]{0.776,0.922,0.757}}0.633 \\
\hline
\end{tabular}
}
\end{table*}

\subsection{Model Compression as a Defense}
% This a task-agnostic methodology.
In Section \ref{subsec:overparams}, we observed a correlation between the model size and the effectiveness of \ourapproach. In this section, we explore to which extent model compression prevents \ourapproach.

\noindent \textbf{Problem Formulation. } Let $f(.)$ be an ML model architecture from a design space $\mathcal{F}$. This model is supposed to be trained on a data distribution $(X,Y) \sim \mathcal{D}^i$ for an original task  $\mathcal{T}_i$. The architecture of $f(.)$ can be described as follows:
\begin{align}
    f(.) = l^{n} \circ l^{n-1} \circ l^{n-2} \circ \dots \circ l^{2} \circ l^{1} s.t. \;\;  f \in \mathcal{F} \label{eqn:ml_layer}\\
    \text{Where for each layer} \quad  l^j  = \{W_1^j, W_2^j, W_3^j, ..., W_{m_j}^j\} \label{eqn:weights}
\end{align}
Where $n$ is the number of neural layers in $f(.)$. Each layer $l^{j}$ is characterized by a specific amount of learnable parameters $W_m^j$; $m$ refers to the layer's width and can be used as a hyperparameter to scale the number of learnable parameters in each layer of $f(.)$. We define our problem of finding a more compact and compressed version $f_{cmp}(.)$ of $f(.)$ as follows:
\begin{equation}
    f_{cmp} = \argmin_{f_m \in \mathcal{F}} \alpha\cdot\mathcal{L}_i(f_{m}(X), Y) + \beta\cdot\mathcal{C}ount(f_{m})  
    \label{eqn:hpo_problem}
\end{equation}
Where $\mathcal{L}_i$ is the loss function used to train $f$ on dataset $D^i$, $ \mathcal{C}ount$ is the function that returns the number of learnable parameters in $f$. $\alpha$ and $\beta$ are control knobs to balance the tradeoff between utility and compression ratio. Here, we abuse the notation and use $m$ as a width scaling factor that can be applied to all the layers of $f(.)$. We apply the same scaling factor to all the layers to preserve the feature abstraction hierarchy along the model's layers. Our objective is to reduce the number of learnable parameters in benign ML models while maintaining the utility of the original task $\mathcal{T}_i$ and lowering the capacity of learning or inferring unrelated features that may be used for hijacking. 

\noindent \textbf{Setup. } We design a search space of $14$ candidate width expansion ratios, ranging from $0.1\times$ to $0.75\times$. For each model $f$, we exhaustively iterate through the search space, sample an expansion ratio $m$, to get a compact model $f_m(.)$. At each iteration, we evaluate $f_m(.)$ using the loss $\mathcal{L}_i(.)$ (as a proxy for utility on the original task) and parameters count $\mathcal{C}ount(.)$ (as a proxy for model size). After the exhaustive search, we rank all candidate $f_{m}(.)$ according to their loss values and parameters count using the TOPSIS method \cite{behzadian2012state} to render an optimal $f_{cmp}(.)$ with a tradeoff between utility and model size. We evaluate our compression method on three hijacking attacks: user re-identification on CK+ \cite{lucey2010extended} identification on Olivetti \cite{samaria1994parameterisation}, and Pneumonia recognition on Chest X-ray \cite{kermany2018identifying}.

\noindent \textbf{Results and Discussion. }
 Results are detailed in Table \ref{tab:comp_results} for each hijacking attack using three models: 2D-CNN, ResNet-9, and MobileNet. Across the $3$ datasets, compact models showed a notable reduction in hijacking accuracy. Specifically, ResNet-9 compact models demonstrate high resilience to hijacking while maintaining comparable accuracy to the baseline model. Identification accuracy on Olivetti and pneumonia type recognition on Chest X-ray decreases by $29\%$ and $38\%$, respectively, even under an aggressive compression strategy with a width expansion of $m=0.1\times$ on the baseline. Conversely, MobileNet shows a modest reduction in hijacking attack vulnerability for re-identification and identification tasks. This finding is consistent with the discussions in Section \ref{subsec:overparams}, highlighting the intrinsic compactness of the baseline MobileNet model. Notably, in the pneumonia scenario, MobileNet with a $m=0.1\times$ shows low hijacking accuracy with a decrease of $11.3\%$ (Logits) and $15.7\%$ (FV). %, stipulating the effectiveness of model compression, even in inherently compact models like MobileNet.

It's worth noting that in specific scenarios, e.g., re-identification on CK+, model compression does not always reduce the hijacking accuracy. This phenomenon may be attributed to the overlap between the original and hijacking tasks, particularly in smaller datasets where models are prone to overfitting. Such models tend to capture fine-grained features necessary for the original task, which concurrently exhibit a high correlation with the hijacking task. Isolating these fine-grained features could compromise the utility of the original task. Therefore, more sophisticated techniques of learning combined with model compression are needed. %, particularly for handling facial-related features in ER systems.

\subsection{Meta-unlearning:}\label{sec:meta-unlearn}

In cases where the hijacking task can be be identified at training time, e.g., biometric identification, we propose to train the model such that it learns the original task and \textit{simultaneously unlearns} the potentially malicious one. The intuition is that some internal representations are more transferable across tasks than others. For example, given an original task $\mathcal{T}_i$ and a related task $\mathcal{T}_j$, i.e., both $\mathcal{T}_i, {T}_j \sim p(\mathcal{T})$, where $p(\mathcal{T})$ is distribution of tasks, a model might learn internal features that are broadly applicable to all tasks in $p(\mathcal{T})$, and others that are specific to the original task $\mathcal{T}_i$. Our objective is to penalize the learning of hijacking task-specific features while learning the ones relevant to the original one. The proposed approach is inspired from the meta-learning literature \cite{maml}: We train the model on $\mathcal{T}_i$, while maximising the loss function $\mathcal{L}_j$, relative to the (hijacking) task $\mathcal{T}_j$. The approach is detailed in Algorithm \ref{alg:maml}. Tables \ref{tab:er_unlearn} and \ref{tab:ukt_unlearn}  show the results of experiments designed to test the effectiveness of the meta-unlearning to retain accuracy on the original classification task under \ourapproach. We also add a hijacking setting with stronger access than our approach, assuming the hijacking dataset can be used to train a neural network for hijacking (NN Surrogate), and not only using distances such as in \ourapproach. Table \ref{tab:ukt_unlearn} shows the results for 'Gender classification' as the original task hijacked for 'Ethnicity recognition'. It shows that when the model is trained with unlearning, the accuracy for the original task generally decreases. This is likely a consequence of the unlearning procedure, which may also remove features that are relevant to the original task. While the surrogate NN achieves, as expected, higher hijacking success, we interestingly notice that the meta-unlearning is more efficient on this attack than \ourapproach, suggesting a higher robustness of distance-based inference in this setting.  For example, the hijacking task accuracy (Ethnicity) has not been significantly impacted by the defense. These results were obtained with $\alpha= 1$ and   $\beta= 0.01$ after empirical exploration. High unlearning coefficient leads to unacceptable accuracy drop in the original task.

\begin{algorithm}[!htp]
\DontPrintSemicolon
  \KwData{$p(\mathcal{T}_i):$ distribution over original task}
  % \KwData{$p(\mathcal{T}_j):$ distribution over sensitive task}
  \KwIn{$\alpha, \beta:$ step size hyperparameters}
  Randomly initialize $\theta$\;
  \While{not done}{
    \texttt{Sample batch of data} $\mathcal{B}_k \sim p(\mathcal{T}_i)$\;
    \For{all $\mathcal{B}_k$}{
      \texttt{Evaluate} $\nabla_{\theta} \mathcal{L}_{\mathcal{T}_i} (f_{\theta})$\;
      \texttt{Update parameters for Original Task:} $\theta'_i = \theta - \alpha \nabla_{\theta} \mathcal{L}_{\mathcal{T}_i} (f_{\theta})$\;
    
      \texttt{Unlearn sensitive task $\mathcal{T}_j$} 
      $\theta \leftarrow \theta + \beta \nabla_{\theta} \mathcal{L}_{\mathcal{T}_j} (f_{\theta'_i})$\;
      }
  }
\texttt{\textbf{Return}}$(f_\theta)$  
\caption{Meta-unlearning }
\label{alg:maml}
\end{algorithm}

\begin{table}[ht]
\centering
\caption{Accuracy of the original task (ER) and hijacking attack (re-identification) with and without meta-unlreaning.}
\label{tab:er_unlearn}
\scalebox{0.98}{
\begin{tabular}{ccccc} 
\hline
\begin{tabular}[c]{@{}c@{}}\textbf{Training }\\\textbf{Strategy}\end{tabular} & \begin{tabular}[c]{@{}c@{}}\textbf{Original }\\\textbf{Model}\\\textbf{(ER)}\end{tabular} & \begin{tabular}[c]{@{}c@{}}\textbf{Original}\\\textbf{Accuracy}\end{tabular} & \begin{tabular}[c]{@{}c@{}}\textbf{\textbf{Hijack by}}\\\textbf{\textbf{NN Surrogate}}\end{tabular} & \begin{tabular}[c]{@{}c@{}}\textbf{\ourapproach~}\\\textbf{(Logits)}\end{tabular} \\ 
\hline
\multirow{3}{*}{\begin{tabular}[c]{@{}c@{}}\textbf{\textbf{Without }}\\\textbf{\textbf{Unlearning}}\end{tabular}} & \textbf{2D-CNN} & {\cellcolor[rgb]{0.984,0.937,0.937}}0.94 & {\cellcolor[rgb]{0.882,0.957,0.906}}0.84 & {\cellcolor[rgb]{0.776,0.922,0.757}}0.29 \\
 & \textbf{ResNet-9} & {\cellcolor[rgb]{0.984,0.937,0.937}}0.95 & {\cellcolor[rgb]{0.882,0.957,0.906}}0.41 & {\cellcolor[rgb]{0.776,0.922,0.757}}0.21 \\
 & \textbf{MobileNet} & {\cellcolor[rgb]{0.984,0.937,0.937}}0.93 & {\cellcolor[rgb]{0.882,0.957,0.906}}0.65 & {\cellcolor[rgb]{0.776,0.922,0.757}}0.24 \\ 
\hline
\multirow{3}{*}{\begin{tabular}[c]{@{}c@{}}\textbf{\textbf{With }}\\\textbf{\textbf{Unlearning}}\end{tabular}} & \textbf{\textbf{2D-CNN}} & {\cellcolor[rgb]{0.976,0.867,0.863}}0.67 & {\cellcolor[rgb]{0.776,0.922,0.757}}0.12 & {\cellcolor[rgb]{0.882,0.957,0.906}}0.33 \\
 & \textbf{\textbf{ResNet-9}} & {\cellcolor[rgb]{0.976,0.867,0.863}}0.62 & {\cellcolor[rgb]{0.776,0.922,0.757}}0.11 & {\cellcolor[rgb]{0.882,0.957,0.906}}0.27 \\
 & \textbf{\textbf{MobileNet}} & {\cellcolor[rgb]{0.976,0.867,0.863}}0.77 & {\cellcolor[rgb]{0.882,0.957,0.906}}0.50 & {\cellcolor[rgb]{0.776,0.922,0.757}}0.23 \\
\hline
\end{tabular}
}
\end{table}

\begin{table}[ht]
\centering
\caption{Accuracy of the original task (Gender) and hijacking attack (Ethnicity) with and without meta-unlreaning.}
\label{tab:ukt_unlearn}
\scalebox{0.98}{
\begin{tabular}{ccccc} 
\hline
\begin{tabular}[c]{@{}c@{}}\textbf{Training }\\\textbf{Strategy}\end{tabular} & \begin{tabular}[c]{@{}c@{}}\textbf{Original }\\\textbf{Model}\\\textbf{(Gender)}\end{tabular} & \begin{tabular}[c]{@{}c@{}}\textbf{Original}\\\textbf{Accuracy}\end{tabular} & \begin{tabular}[c]{@{}c@{}}\textbf{\textbf{Hijack by}}\\\textbf{\textbf{NN Surrogate}}\end{tabular} & \begin{tabular}[c]{@{}c@{}}\textbf{\ourapproach~}\\\textbf{(Logits)}\end{tabular} \\ 
\hline
\multirow{3}{*}{\begin{tabular}[c]{@{}c@{}}\textbf{\textbf{Without }}\\\textbf{\textbf{Unlearning}}\end{tabular}} & \textbf{2D-CNN} & {\cellcolor[rgb]{0.984,0.937,0.937}}0.87 & {\cellcolor[rgb]{0.882,0.957,0.906}}0.51 & {\cellcolor[rgb]{0.776,0.922,0.757}}0.29 \\
 & \textbf{ResNet-9} & {\cellcolor[rgb]{0.984,0.937,0.937}}0.88 & {\cellcolor[rgb]{0.882,0.957,0.906}}0.44 & {\cellcolor[rgb]{0.776,0.922,0.757}}0.28 \\
 & \textbf{MobileNet} & {\cellcolor[rgb]{0.984,0.937,0.937}}0.85 & {\cellcolor[rgb]{0.882,0.957,0.906}}0.46 & {\cellcolor[rgb]{0.776,0.922,0.757}}0.29 \\ 
\hline
\multirow{3}{*}{\begin{tabular}[c]{@{}c@{}}\textbf{\textbf{With }}\\\textbf{\textbf{Unlearning}}\end{tabular}} & \textbf{\textbf{2D-CNN}} & {\cellcolor[rgb]{0.976,0.867,0.863}}0.65 & {\cellcolor[rgb]{0.776,0.922,0.757}}0.15 & {\cellcolor[rgb]{0.882,0.957,0.906}}0.27 \\
 & \textbf{\textbf{ResNet-9}} & {\cellcolor[rgb]{0.976,0.867,0.863}}0.76 & {\cellcolor[rgb]{0.776,0.922,0.757}}0.16 & {\cellcolor[rgb]{0.882,0.957,0.906}}0.32 \\
 & \textbf{\textbf{MobileNet}} & {\cellcolor[rgb]{0.976,0.867,0.863}}0.67 & {\cellcolor[rgb]{0.882,0.957,0.906}}0.45 & {\cellcolor[rgb]{0.776,0.922,0.757}}0.30 \\
\hline
\end{tabular}
}
\end{table}

\begin{table}[ht]
\centering
\caption{Impact of Meta-Unlearning on Original Task and Hijacking Task  Accuracy}
\label{tab:pdd_unlearn}
\scalebox{0.98}{
\begin{tabular}{ccccc} 
\hline
\begin{tabular}[c]{@{}c@{}}\textbf{Training }\\\textbf{Strategy}\end{tabular} & \begin{tabular}[c]{@{}c@{}}\textbf{Original }\\\textbf{Model}\\\textbf{(PDD)}\end{tabular} & \begin{tabular}[c]{@{}c@{}}\textbf{Original}\\\textbf{Accuracy}\end{tabular} & \begin{tabular}[c]{@{}c@{}}\textbf{\textbf{Hijack by}}\\\textbf{\textbf{NN Surrogate}}\end{tabular} & \begin{tabular}[c]{@{}c@{}}\textbf{\ourapproach~}\\\textbf{(Logits)}\end{tabular} \\ 
\hline
\multirow{3}{*}{\begin{tabular}[c]{@{}c@{}}\textbf{\textbf{Without }}\\\textbf{\textbf{Unlearning}}\end{tabular}} & \textbf{2D-CNN} & {\cellcolor[rgb]{0.984,0.937,0.937}}0.803 & {\cellcolor[rgb]{0.882,0.957,0.906}}0.700 & {\cellcolor[rgb]{0.776,0.922,0.757}}0.543 \\
 & \textbf{ResNet-9} & {\cellcolor[rgb]{0.984,0.937,0.937}}0.830 & {\cellcolor[rgb]{0.776,0.922,0.757}}0.648 & {\cellcolor[rgb]{0.776,0.922,0.757}}0.602 \\
 & \textbf{MobileNet} & {\cellcolor[rgb]{0.984,0.937,0.937}}0.793 & {\cellcolor[rgb]{0.882,0.957,0.906}}0.666 & {\cellcolor[rgb]{0.776,0.922,0.757}}0.656 \\ 
\hline
\multirow{3}{*}{\begin{tabular}[c]{@{}c@{}}\textbf{\textbf{With }}\\\textbf{\textbf{Unlearning}}\end{tabular}} & \textbf{\textbf{2D-CNN}} & {\cellcolor[rgb]{0.976,0.867,0.863}}0.625 & {\cellcolor[rgb]{0.776,0.922,0.757}}0.379 & {\cellcolor[rgb]{0.882,0.957,0.906}}0.620 \\
 & \textbf{\textbf{ResNet-9}} & {\cellcolor[rgb]{0.984,0.937,0.937}}0.843 & {\cellcolor[rgb]{0.882,0.957,0.906}}0.651 & {\cellcolor[rgb]{0.882,0.957,0.906}}0.631 \\
 & \textbf{\textbf{MobileNet}} & {\cellcolor[rgb]{0.976,0.867,0.863}}0.753 & {\cellcolor[rgb]{0.776,0.922,0.757}}0.587 & {\cellcolor[rgb]{0.776,0.922,0.757}}0.595 \\
\hline
\end{tabular}
}
\end{table}

% \begin{figure}[!ht]
%   \centering
%   \includegraphics[width=0.95\linewidth]{old_images/2DCNN unlearning ID.pdf}
%   \caption{FER-Trained Model(2D-CNN): Leakage Accuracy in users Re-Identification on Ck+ sub-dataset using ID-Model and distance similarity.}
%   \label{fig:unlearnindID2DCNN}
% \end{figure}

% \begin{itemize}
%     \item \textcolor{blue}{\textbf{Pruning as a defense??}}
%     \item \textbf{what other application cases we want/can show?} -- maybe virtual reality applications? these papers to be checked: \url{https://dl.acm.org/doi/pdf/10.1145/3584931.3607004}; {\url{https://dl.acm.org/doi/10.1145/3460120.3485345}}; \url{https://ieeexplore.ieee.org/abstract/document/10229062};
%     \url{https://arxiv.org/pdf/2301.09041.pdf}
    
%     - Style transfer applications? \url{https://openaccess.thecvf.com/content/CVPR2022/papers/Hu_Protecting_Facial_Privacy_Generating_Adversarial_Identity_Masks_via_Style-Robust_Makeup_CVPR_2022_paper.pdf}

%     - Deepfake? \url{https://openaccess.thecvf.com/content/WACV2023/html/Ciftci_My_Face_My_Choice_Privacy_Enhancing_Deepfakes_for_Social_Media_WACV_2023_paper.html}

%     \item \textcolor{blue}{\textbf{VR case}}: \url{https://www.frontiersin.org/articles/10.3389/fpsyg.2022.864266/full}

%     \item \textcolor{blue}{\textbf{ER in the VR-- code needs to be requested}}:\url{https://link.springer.com/article/10.1007/s10055-022-00720-9}
% \end{itemize}

\section{Discussion} \label{sec:disc}

\noindent \textbf{Potential for harm with less access privilege.}  In this paper, we introduce a novel threat model within the context of model hijacking attacks. Specifically, we identify a new risk where adversaries with very restricted access can hijack ML models. Unlike existing hijacking attacks, which are constrained by the number of classes in the original task, our approach imposes no such limitations on the hijacking task's class count. Consequently, \ourapproach~ not only assumes a more robust threat model but also enables attackers with lower access privileges to inflict greater harm. Moreover, \ourapproach~ does not require input modification during inference, facilitating a more covert form of hijacking. This allows a malicious model owner to repurpose the model while ostensibly complying with AI regulations, as the attack can be executed using a model from a trusted vendor without altering the input data.

%justifying compliance with the trusted vendor, and no interference with the input. 

% \noindent \textbf{Insights towards security-aware NAS.}  We suggest that the vulnerability of ML models to this type of attack is primarily related to over-parametrization. Interestingly, our results reveal a disparity in attack success based on the model architecture. This observation could be particularly significant in Neural Architecture Search towards developing secure-by-design models.
% \noindent \textbf{Ethical considerations} 

\noindent \textbf{Beyond the technical implications.}  We contend that our work has also significant implications for risk-based regulatory frameworks, as it challenges some of their foundational assumptions. In fact, the debate on regulating AI-powered systems has gained global momentum ultimately exemplified by initiatives like the European Commission's proposal for a risk-assessment framework known as the EU AI-Act \cite{aiact}. This framework seeks to categorize potential risks based on the nature of the task for which the ML model is trained, i.e., the original task. The reliance on the learned task's criticality as a risk metric is inevitably tied to the following implicit hypothesis: 
"If an ML model is trained on a data distribution $\mathcal{D}$, to learning a task $\mathcal{T}$, it is unlikely to (unintentionally) learn another task $\mathcal{T'}$". Our work shows that it is not sufficient for a benign model to be securely trained on a legitimate original task to guarantee that it will be immune to repurposing for unethical or illegitimate tasks. Therefore, additional safeguards and assurances are necessary.
%We believe \ourapproach~ challenges some of the AI regulatory efforts' foundations and underscores the need for more comprehensive approaches.
%The potential for compliant ML models to be repurposed for unethical or illegal tasks poses significant accountability risks for model owners. This highlights a critical oversight in the EU AI Act and similar regulatory frameworks, which assume that models trained on specific tasks are confined to those boundaries.

%We contend that our work has also significant implications for risk-based regulatory frameworks, as it challenges some of their foundational assumptions.

\noindent \textbf{Attack Limitations.} In our study, we demonstrate that the hijacking effectiveness of \ourapproach~can reach state-of-the-art task accuracy in some cases. However, this effectiveness can be influenced by several factors, particularly the victim model's architecture. Because of its zero-shot aspect, the performance of SnatchML can be significantly affected by the distribution of the hijacking task, (inter-class/intra-class variance). For instance, in scenarios like ethnicity classification, different ethnic groups might have significant intra-class variance. Despite this limitation, our approach represents a significant advancement over traditional hijacking methods by overcoming the constraints related to the number of classes in the hijacking task. 

\section{Related work}\label{sec:related}
%\ourapproach~falls within the model hijacking literature. We distinguish 2 main categories of related work: 

\noindent\textbf{Model hijacking.} Model hijacking has emerged as a critical concern in the deployment of ML systems. Traditional hijacking techniques typically necessitate access to the model's training process, allowing for the insertion of malicious alterations during training to facilitate the hijacking. 
Several papers proposed to overload ML models with secondary tasks. Salem et al.~\cite{salem2021get} proposed Model Hijacking attack that hides a model covertly while training a victim model. Mallya et al.~\cite{mallya2018packnet} proposed Packnet that trains the model with multiple tasks. Si et al.~\cite{si2023two}, propose the Ditto attack, which hijacks text generation models by camouflaging hijacking datasets to resemble original ones, focusing on the output rather than input, making the hijack difficult to detect. The authors in \cite{salem2021get} introduce the Chameleon and Adverse Chameleon attacks, which utilize an encoder-decoder model to camouflage hijacking datasets to appear like the original dataset, allowing stealthy hijacking of machine learning models.

% These methods often rely on attack settings that are similar to poisoning/backdoor attacks. 
The main difference with our work is that these attacks assume a threat model similar to poisoning attacks with the attacker having access to the training data/process, while \ourapproach~ operated fully at inference time. 
Song et al. introduced the notion of overlearning \cite{song2019overlearning}, proposing its use to perform repurposing attacks. Their method involves training a separate adversary model to infer sensitive attributes from a trained model’s representations. This adversary model is trained on labeled data to predict sensitive attributes based on extracted representations. Additionally, \cite{song2019overlearning} demonstrates how a trained model can be reused to predict a different sensitive attribute by fine-tuning a new classifier added to an existing model layer.
In contrast to \ourapproach~, \cite{song2019overlearning} requires additional training and labeled data, making it resource-intensive. Moreover, their approach operates in a white-box setting, whereas SnatchML can function effectively in a black-box scenario.

\textbf{Transfer Learning and Adversarial Reprogramming}:
Transfer learning (TL) approaches involve fine-tuning a model fully or partially to adapt to a new domain, leveraging the knowledge gained from the source domain \cite{weiss2016survey}. SnatchML differs from standard TL in that it does not require model fine-tuning.  This is especially beneficial in black-box scenarios, such as cloud infrastructures, where attackers do not have access to gradients or internal model components. SnatchML operates in cases where transfer learning is not possible, allowing attackers to abuse model outputs without directly interacting with the model.

AR operates at test time; repurposes a pretrained neural network for a new task by applying a specific transformation to the input data, effectively mapping it from the target domain to the source domain \cite{elsayed2018adversarial}. This transformation, or adversarial program, is optimized so that the network's output for the transformed inputs aligns with the desired outputs of the new task. Specifically, the input transformation and output mapping are designed to replace the transfer learning, effectively hijacking the model's capabilities by manipulating the input and output but without altering its parameters. 
Some AR approaches include training a neural network for the output mapping. For example, in \cite{trainableAR, AR_cross} the approach consists of adding a trainable dense layer between the source model’s output  and the target model’s output, which shifts AR closer to a transfer-learning paradigm.

Our approach has several key distinctions with AR:
First, similar to other hijacking attacks, AR is constrained by the number of classes present in the victim model, as it repurposes the model’s existing outputs for the new task, unless a TL approach is used. In contrast, \ourapproach handles a higher number of classes for the hijacking task, offering greater flexibility.
Second, AR typically requires white-box access to the model to leverage gradients for optimization. This limits its applicability in settings where only black-box access is available. SnatchML, however, can be executed effectively in both white-box and black-box settings.
Finally, AR  requires modifying the input, which can be impractical when the attacker, who is also the model owner, aims to covertly misuse a benign model for malicious purposes. In such scenarios, altering the model or its inputs could raise suspicion, rendering AR unsuitable. In contrast, our approach operates without any modifications to the input or the model, enabling more discreet exploitation. We include a comparative analysis with AR in Appendix \ref{sec:reprog}. % (Fig. \ref{fig:compare_adv_reprogram}).

\section{Conclusion}
This paper introduces SnatchML, a novel framework for hijacking ML models at inference time, demonstrating a novel threat model for model hijacking. By exploiting the output of the victim model during inference, SnatchML effectively repurposes models for unintended tasks without requiring access to training data and without reprogramming in the input space. Our experiments reveal that SnatchML can perform hijacking tasks even when they are unrelated to the original model's purpose, highlighting the increased risk and potential for harm with limited access privileges. Additionally, we investigate the role of over-parameterization in facilitating these attacks.



\bibliographystyle{IEEEtranS}
\bibliography{main}


% \end{thebibliography}

%\newpage
%\clearpage
\newpage
\appendix

\section{Appendix}
\label{sec:app1}

\subsection{Comparison with Adversarial Reprogramming}\label{sec:reprog}
\begin{figure*}[!h]
  \centering
  \includegraphics[width=\linewidth]{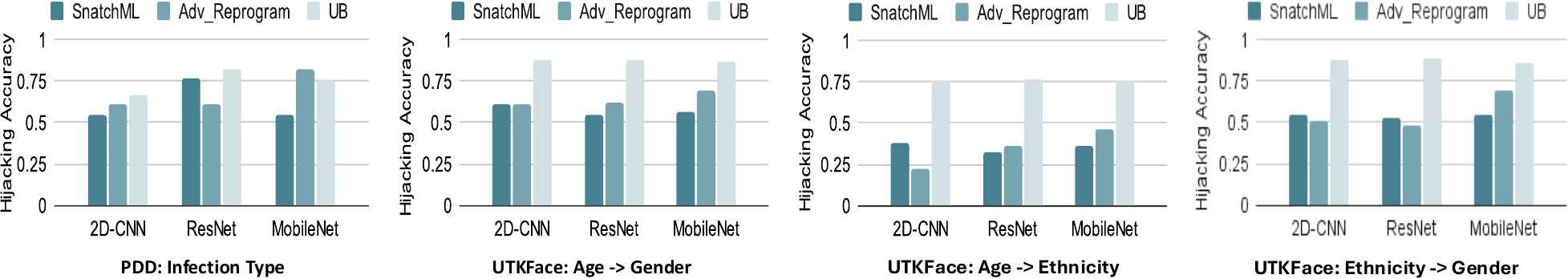}
  \caption{Comparison between the hijacking attack performance with SnatchML, Adversarial Reprogramming \cite{elsayed2018adversarial}, and Unconstrained upper bound \cite{salem2021get}. }
  \label{fig:compare_adv_reprogram}
\end{figure*}

Figure \ref{fig:compare_adv_reprogram} shows the results of \ourapproach~  comparatively with adversarial reprogramming. To  be able to establish a comparison with adversarial reprogramming, we only provide results on the feasible combinations of original-hijacking tasks, i.e., where the number of hijacking task classes is less than the number of original task classes.

\begin{figure*}[htp]
  \centering
  \includegraphics[width=\linewidth]{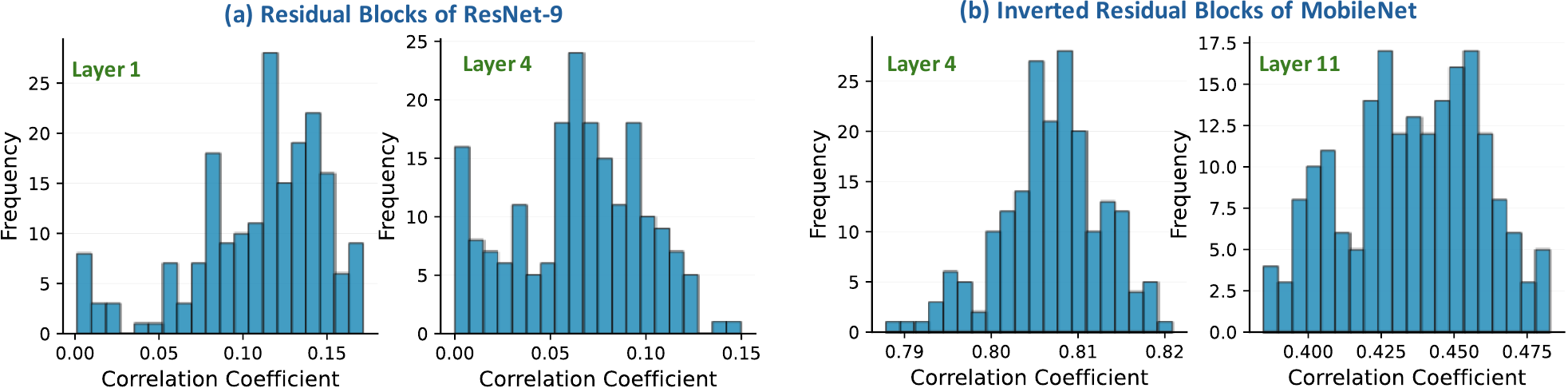}
  \caption{Correlation distribution between output feature maps from pretraine Emotion Recognition and Face Recognition.}
  \label{fig:corr3}
\end{figure*}

\subsection{Features Correlation}\label{sec:intuition}

 We posit that the models might learn more than they should in the training process, including learning other tasks. Our intuition is that ML models, perhaps due to over-parametrization, have a by-design capacity that exceeds the minimum necessary capacity to learn the task they are trained for. While it has been shown in the literature that models can overfit and memorise data making them vulnerable to certain attacks such as Membership \cite{reza} and property inference \cite{pia} attacks, the risk that ML models (unintentionally) generalize to other tasks is yet to be explored/exploited.
In this section, we provide a preliminary analysis to gain insights in the plausibility of this hypothesis. Given a model $f^{(i)}(.)$ trained on a data distribution $(X,Y) \sim \mathcal{D}$, using a loss function $\mathcal{L}_i$, relevant to a Task $\mathcal{T}_i$. Let $f^{(j)}(.)$ be a model with the same architecture as $f^{(i)}(.)$, but is trained  on a data distribution $(X’,Y’) \sim \mathcal{D}’$, using a loss function $\mathcal{L}_j$, relevant to a Task $\mathcal{T}_j \neq \mathcal{T}_i$. We ask whether
features learnt by $f^{(i)}(.)$ are correlated with those learnt by $f^{(j)}(.)$? 
We train two ResNet-9 and two Mobilenet architectures each for both Emotion Recognition and Face Recognition. We then check the correlation coefficient distribution between the same layers’ features maps that correspond to the 2 tasks.  
The correlation coefficient between two variables $X$ and $Y$, denoted as $r$, is defined as:
\begin{equation}
    r = \frac{\sum_{i=1}^n (X_i - \overline{X})(Y_i - \overline{Y})}{\sqrt{\sum_{i=1}^n (X_i - \overline{X})^2} \sqrt{\sum_{i=1}^n (Y_i - \overline{Y})^2}}
\end{equation}
where:
\begin{itemize}
    \item $n$ is the number of data points,
    \item $X_i$ and $Y_i$ are the individual values of the variables $X$ and $Y$,
    \item $\overline{X}$ and $\overline{Y}$ are the means of $X$ and $Y$ respectively.
\end{itemize}

% The value of $r$ ranges from -1 to 1, where:
% \begin{itemize}
%     \item $r = 1$ indicates a perfect positive linear relationship,
%     \item $r = -1$ indicates a perfect negative linear relationship,
%     \item $r = 0$ indicates no linear relationship between the variables.
% \end{itemize}

Figure \ref{fig:corr3} shows the features correlation distribution for Layer 5 and layer 11 for ResNet-9, and Layer 2 and Layer 4 for Mobilenet. Intrestingly, Figure \ref{fig:corr3} shows a positive correlation suggesting a potential overlap between what the model learns for tasks that share semantic clues. In other words, this preliminary observation supports the idea that models trained on an original task are capable of extracting features relevant to other tasks.

\subsection{Impact of limited output classes}
In continuation of our study on the performance of the SnatchML attack, we now explore how varying the number of top logits impacts the effectiveness of repurposing a pre-trained model for a new classification task. This analysis investigates the use of different subsets of the highest-ranked logit values from a pre-trained model originally intended for a different task. The motivation behind this study lies in understanding how the selection of top-ranked logit values influences the effectiveness of the SnatchML attack. By ranking and selecting the top k logits from the model's outputs, we investigate how accurately the model can be hijacked for a new classification task.

We illustrate here three distinct cases where a pretrained MobileNet model, originally trained on ImageNet, is adapted for ethnic classification using the UTKface dataset, pneumonia classification using chest X-ray images and face recognition using Olivetti dataset. Our investigation focuses on how different levels of high-confidence predictions (top-k logits) influence the effectiveness of the SnatchML attack.
By varying the parameter k (representing the number of top-ranked logits used), we explore the model's vulnerability to repurposing for these new tasks, which were not part of its original training scope. The experimental results are presented across different $k$ values—specifically, $k=1$,$10$ ,$100$ , and $1000$. For each $k$, the accuracy metrics reveal the model's performance in correctly predicting ethnicity and pneumonia.

\begin{figure*}[htp]
  \centering
  \includegraphics[width=\linewidth]{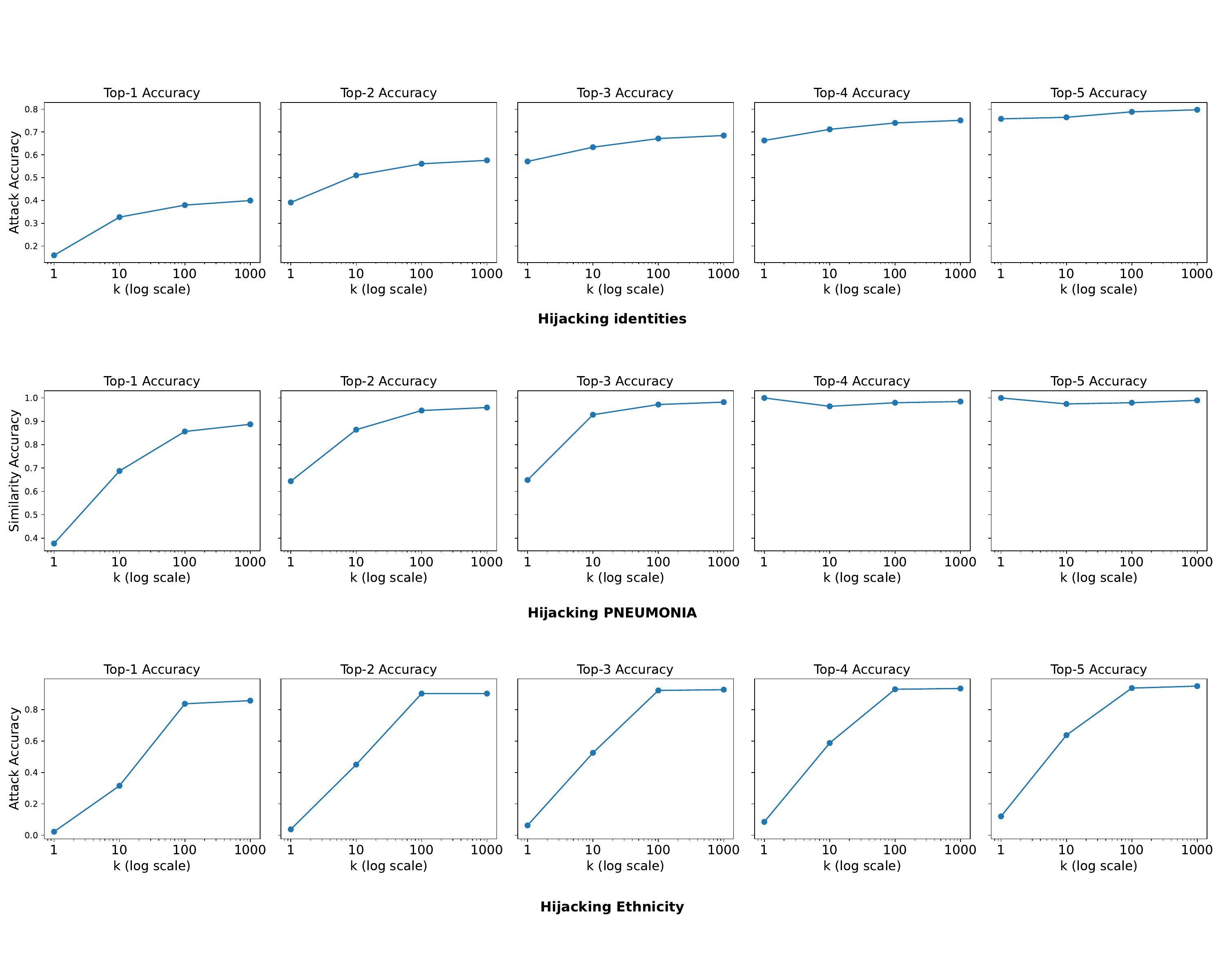}
  \caption{"Effectiveness of SnatchML Attack: Impact of Top Logits Selection Across Diverse Hijacking Tasks.}
  \label{fig:klogit}
\end{figure*}

Across the three datasets, the results, presented in Figure \ref{fig:klogit} exhibit a consistent trend: as k increases, the model's accuracy improves. This trend indicates that the critical information required for the SnatchML attack—utilizing the model's ability to classify ethnicities, pneumonia and faces is concentrated within the top logits of its prediction outputs. Notably, there is a noticeable convergence in accuracy metrics as k increases, particularly evident from k=100 onwards. This convergence suggests that a significant portion of the useful information for our attack is captured within the top $10\%$ of the highest-ranked logits. This finding underscores that a relatively small subset of confident predictions suffices to effectively repurpose the model for these unintended classification tasks. Furthermore, additional experiments confirming similar observations are presented in the appendix.

\subsection{Impact of the hijacking task's complexity}
\begin{figure}
   \centering
  \includegraphics[width=\linewidth]{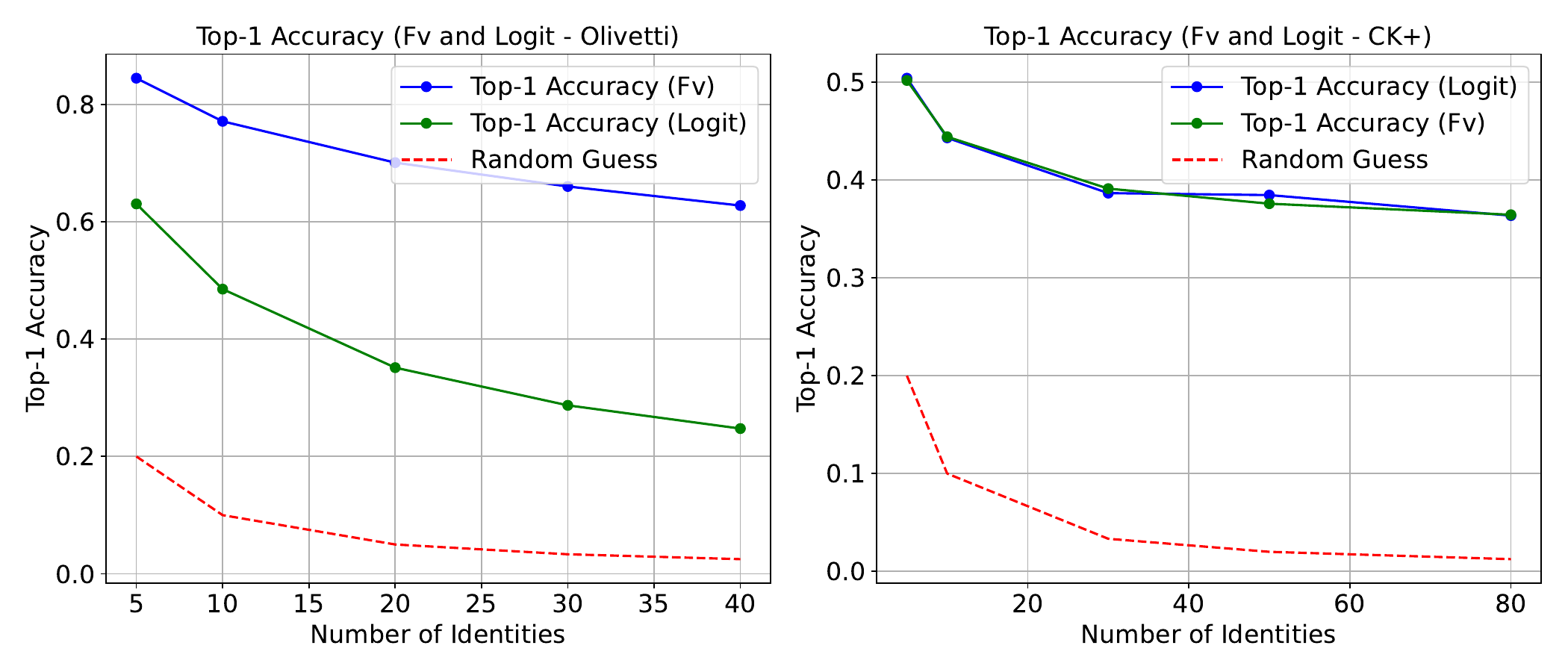}
  \caption{Exploring the Relationship Between Complexity (Identity Count) and Attack Efficiency (Top-1 Accuracy.) }
  \label{fig:Complexity}   
\end{figure}

To learn more about how task complexity affects attack performance, we ran an additional experiment in both the Olivetti and CK+ datasets, varying the number of identities. This enabled us to analyze how the SnatchML assault behaved as the secondary task's complexity rises, offering more insight into the relationship between task difficulty and attack efficacy. The Figure \ref{fig:Complexity} represents the relationship between Top-1 accuracy and the number of identities for the two datasets, highlighting the impact of dataset complexity on model performance in a controlled threat model scenario with a ResNet-18 model trained for ER. Top-1 accuracy for both datasets drops as the number of identities increases, which is to be expected given that the task gets fundamentally more difficult with more identities. The Olivetti dataset consistently exhibits higher Top-1 accuracy than CK+, indicating that the identities in Olivetti are easier for the model to distinguish. This difference can be attributed to the greater similarity between classes in CK+, where subtle variations make accurate predictions more difficult. The intraclass/interclass variation ratios support this observation; Olivetti’s ratio of 0.65 suggests relatively distinct identities, while CK+’s higher ratio of 0.88 indicates that intraclass variation is closer to interclass variation, reflecting increased complexity. Within the black-box, inference-only threat model, where the attacker lacks direct input access, this analysis helps explain why accuracy is particularly low for complex datasets like CK+. The results demonstrate that the subtle differences between classes in datasets with higher complexity create significant challenges for the model, especially when direct access to input manipulation or gradient feedback is unavailable.

%\textcolor{blue}{\hl{this could go to the appendices -- it is no longer important in the current frame of the paper -- \\}}

% #\begin{figure}[hp]
% #  \centering
% #  \includegraphics[width=1\linewidth]#{old_images/pathology_distribution.pdf}
% #  \caption{Pneumonia confusion matrix #\textcolor{blue}{\hl{this should move #to the apendix--}}}
% #  \label{fig:pneumonia_proprety_attack}
% #\end{figure}

% \subsection{property inference}

% \begin{figure}[hp]
%   \centering
%   \includegraphics[width=1\linewidth]{old_images/matrix_2dcnn_gender_distribution.pdf}
%   \caption{Confusion matrix ethnic to gender Property inference attacks \textcolor{red}{HB: Too small, what about only showing predicted/real distribution?}}
%   \label{fig:matrix ethnic to gender}
% \end{figure}

% #\begin{figure}[t]
% #  \centering
% #  \includegraphics[width=1\linewidth]{old_images/matrix_2dcnn_ethnic_to_age-2.pdf}
% #  \caption{Confusion matrix age #Property inference attacks #\textcolor{red}{HB: Too small, what #about only showing predicted/real #distribution?}}
% #  \label{fig:matrix ethnic to age}

\subsection{Model Hijacking for Unrelated Tasks} 

To investigate the generalizability of model hijacking to tasks unrelated to the original training task, we conducted experiments using ResNet-18, MobileNet, and ResNet-50. These models, originally trained on ImageNet or CIFAR-10, were evaluated on a variety of unrelated tasks, including MNIST, CIFAR-100, CK+ (emotion and identity recognition), Olivetti, the Celebrity dataset, and a synthetic dataset. Results are shown in Tables \ref{tab:resnet-18-imagenet}, \ref{tab:mobilenet-imagenet}, \ref{tab:Resnet_50_cifar10} and \ref{tab:mobilenet_cifar10}.

% Table for ResNet-18
\begin{table}[h]
    \centering
    \caption{Hijacking attack on ResNet-18 Trained on Unrelated Tasks (ImageNet) Across Multiple Applications}
    \label{tab:resnet-18-imagenet}
    \scalebox{0.83}{
    \begin{tabular}{lcccc}
        \hline
        \textbf{Application} & \textbf{Random Guess} & \textbf{Trained on ImageNet (Fv)} & \textbf{Trained on ImageNet (logit)} \\ \hline
        MNIST & 10\% & 93\% & 91\% \\
        CIFAR100 & 1\% & 23\% & 20\% \\
        CK+ emotions & 16.6\% & 56\% & 60\% \\
        CK+ identities & 1.1\% & 65\% & 63\% \\
        Olivetti & 2.5\% & 91\% & 87\% \\
        Celebrity & 11\% & 100\% & 100\% \\
        Synthetic & 2.1\% & 100\% & 100\% \\ \hline
    \end{tabular}
    }
\end{table}

% Table for MobileNet
\begin{table}[h]
    \centering
    \caption{Hijacking attack on MobileNet Trained on Unrelated Tasks (ImageNet) Across Multiple Applications}
        \label{tab:mobilenet-imagenet}

        \scalebox{0.83}{
    \begin{tabular}{lcccc}
        \hline
        \textbf{Application} & \textbf{Random Guess} & \textbf{Trained on ImageNet (Fv)} & \textbf{Trained on ImageNet (logit)} \\ \hline
        MNIST & 10\% & 86\% & 85\% \\
        CIFAR100 & 1\% & 15\% & 14\% \\
        CK+ emotions & 16.6\% & 48\% & 53\% \\
        CK+ identities & 1.1\% & 68\% & 57\% \\
        Olivetti & 2.5\% & 76\% & 73\% \\
        Celebrity & 11\% & 100\% & 100\% \\
        Synthetic & 2.1\% & 100\% & 100\% \\ \hline
    \end{tabular}
    }
\end{table}

% Table for ResNet-50
\begin{table}[h]
    \centering
    \caption{Hijacking attack on ResNet-50 Trained on Unrelated Tasks (Cifar-10) Across Multiple Applications}
        \label{tab:Resnet_50_cifar10}
        \scalebox{0.83}{
    \begin{tabular}{lcccc}
        \hline
        \textbf{Application} & \textbf{Random Guess} & \textbf{Trained on CIFAR-10 (Fv)} & \textbf{Trained on CIFAR-10 (logit)} \\ \hline
        MNIST & 10\% & 91\% & 62\% \\
        CIFAR100 & 1\% & 13\% & 6\% \\
        CK+ emotions & 16.6\% & 48\% & 48\% \\
        CK+ identities & 1.1\% & 65\% & 45\% \\
        Olivetti & 2.5\% & 94\% & 64\% \\
        Celebrity & 11\% & 100\% & 100\% \\
        Synthetic & 2.1\% & 99\% & 99\% \\ \hline
    \end{tabular}
    }
\end{table}

% Table for MobileNet on CIFAR-10
\begin{table}[h]
    \centering
    \caption{Hijacking attack on MobileNet Trained on Unrelated Tasks (Cifar-10) Across Multiple Applications}
            \label{tab:mobilenet_cifar10}
        \scalebox{0.83}{
    \begin{tabular}{lcccc}
        \hline
        \textbf{Application} & \textbf{Random Guess} & \textbf{Trained on CIFAR-10 (Fv)} & \textbf{Trained on CIFAR-10 (logit)} \\ \hline
        MNIST & 10\% & 58\% & 57\% \\
        CIFAR100 & 1\% & 7\% & 6\% \\
        CK+ emotions & 16.6\% & 44\% & 51\% \\
        CK+ identities & 1.1\% & 49\% & 49\% \\
        Olivetti & 2.5\% & 49\% & 43\% \\
        Celebrity & 11\% & 100\% & 100\% \\
        Synthetic & 2.1\% & 99\% & 99\% \\ \hline
    \end{tabular}
    }
\end{table}

% We aimed to assess whether the model could be effectively hijacked for a task, such as MNIST digit classification, without any prior training on that specific task. To illustrate this, we generated t-SNE plots comparing the feature distributions of MNIST digit classes using the feature vectors extracted from the pretrained ResNet18 model and a randomly initialized ResNet18 model (see Figure \ref{fig:mnist-tsne}. Surprisingly, we observed distinct clusters corresponding to different digit classes in both cases despite neither model being trained on the MNIST dataset. This observation underscores the potential for model hijacking even when the hijacking task is entirely unrelated to the original task. The presence of discernible clusters in the feature space, as depicted in the figure \ref{fig:mnist-tsne}, serves as compelling evidence to further explore the vulnerabilities of pretrained models to hijacking attacks.

% \begin{figure}[htp]
%   \centering
%   \includegraphics[width=0.95\linewidth]{attack_figures/tsne_mnist.png}
%   \caption{Comparison of t-SNE distributions for MNIST digit classes using ResNet18 pretrained on ImageNet (left) and randomly initialized ResNet18 (right), based on their feature vectors.}
%   \label{fig:mnist-tsne}
% \end{figure}

\begin{figure*}[htp]
  \centering
  \includegraphics[width=\linewidth]{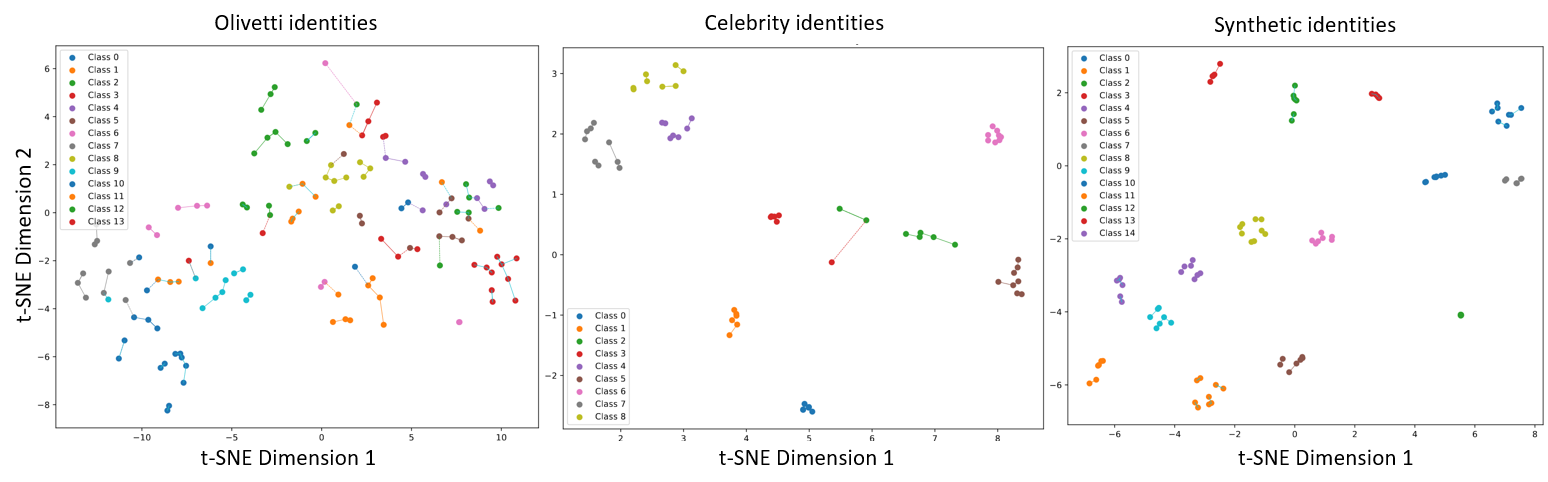}
  \caption{t-SNE distributions for identity classes across three datasets (Olivetti, Celebrity, and Synthetic) using ResNet18 pretrained on ImageNet, based on their feature vectors. The Euclidean distance illustrates the closest data points: solid lines connect data points of the same identity, while dashed lines connect data points of different identities.}
  \label{fig:tsne_distrib}
\end{figure*}

To further illustrate our finding, Figure \ref{fig:tsne_distrib} illustrates the t-SNE distribution of identity classes for three datasets (Olivetti, Celebrity, and Synthetic) based on their feature vectors derived from inference using a ResNet18 model pretrained on ImageNet. The figures distinctly exhibit the existence of separate clusters corresponding to each identity. This finding further reinforces the notion of model hijacking's potential, as it demonstrates that pretrained models can capture meaningful representations even for tasks unrelated to their original training, echoing the observations made in the preceding paragraph.

Table \ref{tab:2d_cnn_random} presents the results of additional experiments evaluating the performance of randomly initialized neural networks across various applications, further illustrating the capabilities of models in hijacking tasks that are entirely unrelated to their original training.

% Table for 2D-CNN Performance
\begin{table}[h]
    \centering
    \caption{Hijacking attack on a randomly initialized 2D-CNN model to perform ER, Identification on CK+ and image classification on MNIST and CIFAR100}
    \label{tab:2d_cnn_random}
    \scalebox{0.83}{
    \begin{tabular}{lccc}
        \hline
        \textbf{Application} & \textbf{Random Guess} & \textbf{Trained on CK+ (Fv)} & \textbf{Trained on CK+ (logit)} \\ \hline
        Emotion (CK+) & 16.6\% & 48\% & 42\% \\
        Identity (CK+) & 1.1\% & 67\% & 47\% \\
        Cifar100 & 1\% & 5\% & 2.3\% \\
        MNIST & 10\% & 80\% & 39\% \\ \hline
    \end{tabular}
    }
\end{table}

% \label{apx:reid_training}

\subsection{Assessing Model Hijacking Across Additional Qualitative Results}
\begin{figure}[htp]
   \centering
  \includegraphics[width=\linewidth]{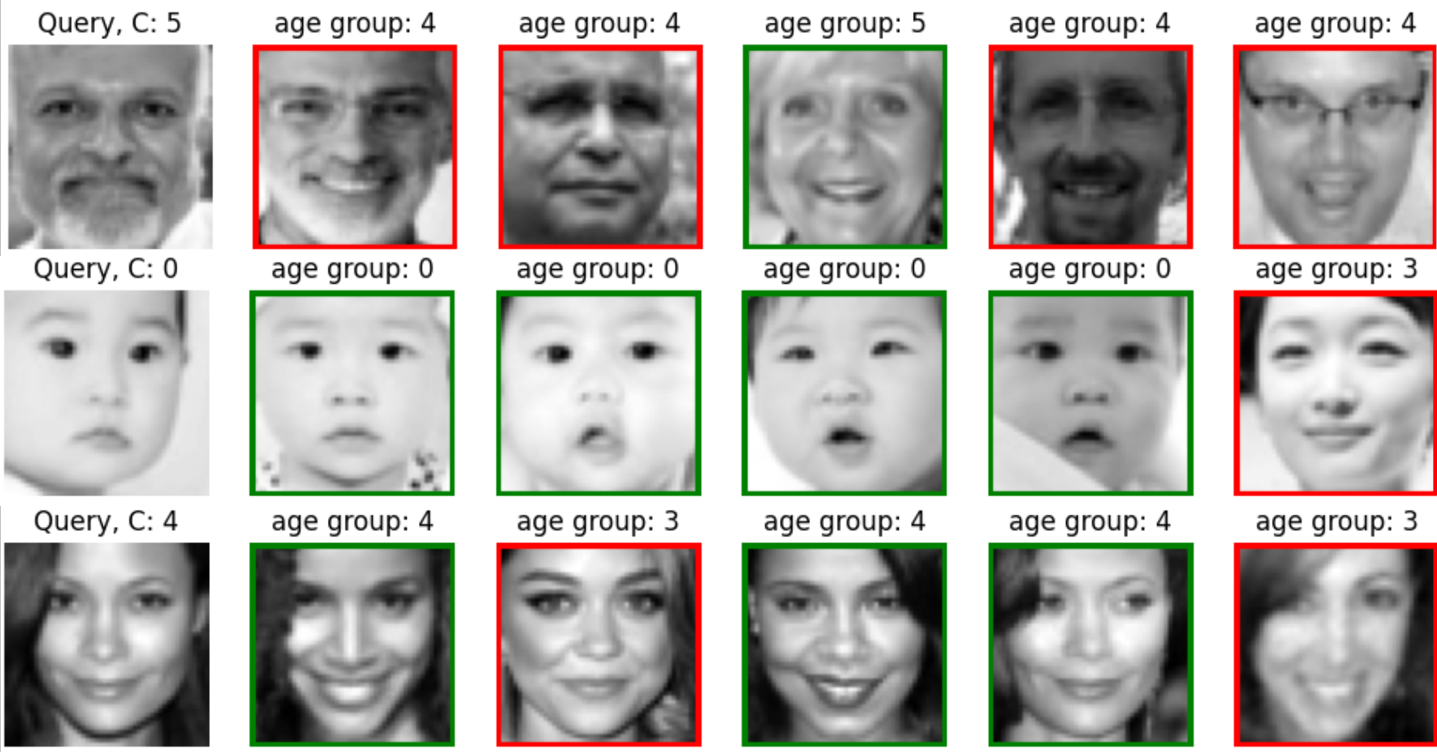}
  \caption{Ethnic groups trained model used to predict age group on UTKface dataset. The 5 nearest images are displayed for each query.}
  \label{fig:to5_ethny_age}   
\end{figure}

\begin{figure}[htp]
  \centering
  \includegraphics[width=\linewidth]{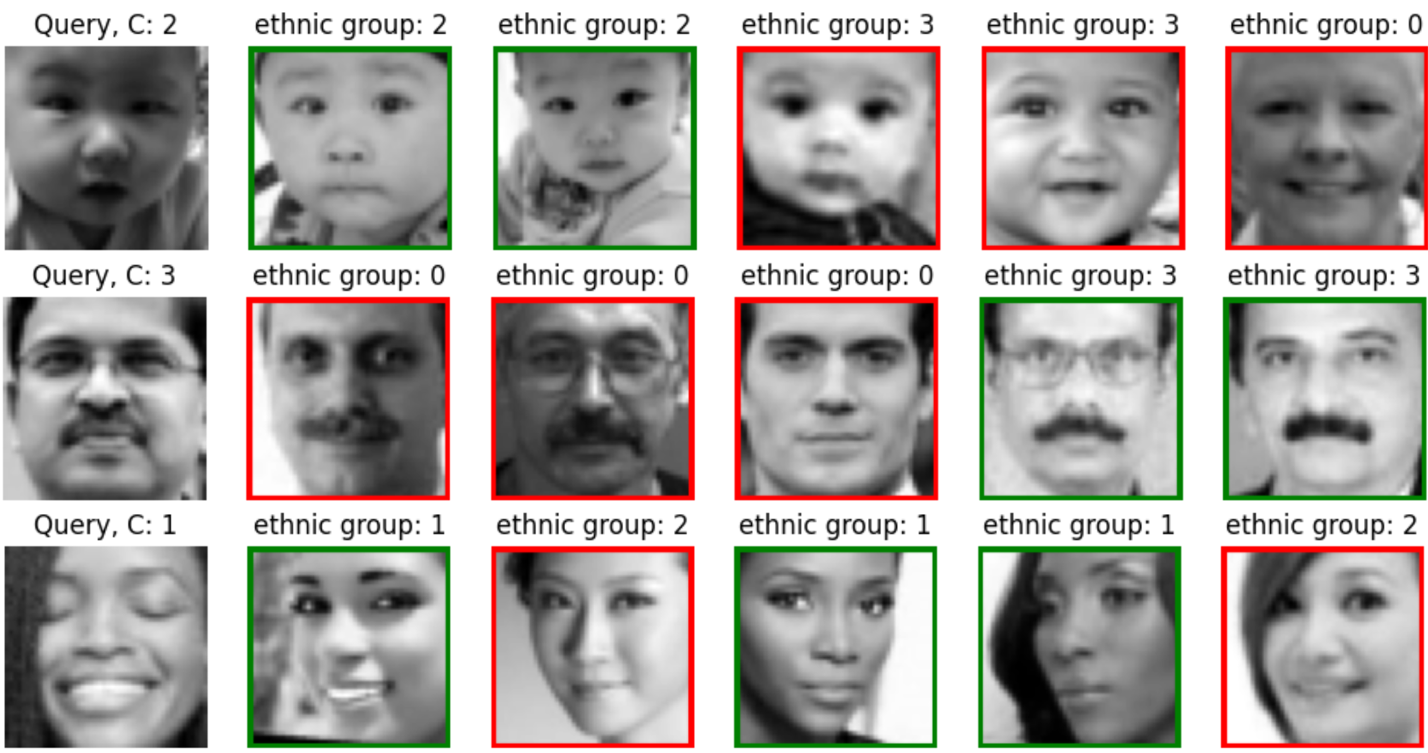}
  \caption{Gender groups trained model used to predict ethnic group on UTKface dataset. The 5 nearest images are displayed for each query.}
  \label{fig:to5_gender_age}
\end{figure}

In this section, we present additional qualitative results for the scenario discussed in Section \ref{sec:scenario5_age}. In Figure \ref{fig:to5_ethny_age}, we depict some examples of predictions made by an Ethnic Groups trained model used for predicting age groups on the UTKface dataset. The five nearest images are displayed for each query. For instance, in the first row, despite the fact that the two closest images do not belong to the correct age group (old), they are both individuals within the age group boundary (adult). Additionally, we observe in the third row that both errors belong to the nearest age group, which could be considered as containing relatively useful information. In Figure \ref{fig:to5_gender_age}, we present some examples of results obtained from a Gender Groups trained model used for predicting ethnic groups on the UTKface dataset. The five nearest images are displayed for each query. In the second row, we observe that the model returns images with similar attributes such as mustaches or glasses, which are unrelated to the original task it was trained for, namely predicting Gender groups. This result illustrates the fact that the model unintentionally learns additional information that can be exploited in unauthorized ways. \vspace{-8pt}

\end{document}